\documentclass[reprint,amsmath,amssymb,aps,prd,twoside]{revtex4-2}
\usepackage{graphicx}
\usepackage{braket}
\usepackage{bm}
\usepackage{comment}
\usepackage{here}
\usepackage{mathtools}
\usepackage[utf8]{inputenc}
\usepackage{booktabs}
\usepackage{slashed}
\usepackage{tikz}
\usepackage{tikz-feynman}
\usepackage{mathrsfs}
\usepackage{slashed}
\usepackage[fit]{truncate}
\usepackage[normalem]{ulem}

\providecommand{\eprint}[2][]{\url{#2}}

\usetikzlibrary{arrows.meta,decorations.markings,calc}
\tikzset{
fermion/.style={
    line width=0.9pt,
    postaction={decorate},
    decoration={
        markings,
        mark=at position 0.55 with {\arrow{Stealth[length=2mm]}}
    }
},
scalar/.style={
    dashed,
    line width=0.9pt
},
cross/.style={
    line width=0.8pt
},
 ferm/.style={
   line width=0.8pt,
   postaction={decorate},
   decoration={markings,mark=at position .58 with {\arrow{Stealth[length=1.8mm]}}}
 },
 scalar mid/.style={
   dashed,
   line width=.8pt,
   postaction={decorate},
   decoration={markings,mark=at position .54 with {\arrow{Stealth[length=1.6mm]}}}
 },
 plain/.style={line width=0.8pt},
 mass/.style={line width=.8pt}
}

\makeatletter
\def\ps@revtexheaders{
  \def\@oddhead{
    \vbox{
      \hbox to \textwidth{
        \textsc{\truncate{11.5cm}{Leptogenesis via Resonant Sequential Dominance and TBC3 Mixing in a Type-I Seesaw Model}}\hfill 
        PHYS. REV. D \textbf{xxx}, xxxx (2026)            
      }
      \vskip 4pt
      \hrule height 0.5pt width \textwidth                   
    }
  }
  \def\@evenhead{
    \vbox{
      \hbox to \textwidth{
        \textsc{Michael Fodroci} and \textsc{Teruyuki Kitabayashi}\hfill 
        XXXXX XXXX X \textbf{xxx}, xxx (2026)
      }
      \vskip 4pt
      \hrule height 0.5pt width \textwidth                   
    }
  }
}
\makeatother

\begin{document}

\title{Leptogenesis via Resonant Sequential Dominance and TBC3 Mixing \\ in a Type-I Seesaw Model}

\author{Michael Fodroci}\thanks{5mtad005@tokai.ac.jp}
\affiliation{Graduate School of Science and Technology, Tokai University, \\Micro/Nano Technology Center, Tokai University, 4-1-1 Kitakaname, Hiratsuka, Kanagawa 259-1292, Japan} 

\author{Teruyuki Kitabayashi}\thanks{teruyuki@tokai.ac.jp}
\affiliation{Department of Physics, School of Science, Tokai University, 4-1-1 Kitakaname, Hiratsuka, Kanagawa 259-1292, Japan} 

\date{April 2026}

\begin{abstract}
We demonstrate how phenomenological model construction assuming Sequential Dominance can be extended to cases with degenerate right-handed neutrino masses. Because this framework accommodates resonant leptogenesis, we designate it as Resonant Sequential Dominance. We then investigate Dirac mass matrix textures compatible with current neutrino data within Resonant Sequential Dominance, utilizing a novel neutrino mixing matrix ansatz termed TBC3, proposed in this work. Additionally, we present an $A_4 \times (Z_3)^4 \times (Z_2)^2$-symmetric Lagrangian that is consistent with neutrino oscillation data and successfully drives resonant leptogenesis. Lastly, we find that accounting for the observed baryon asymmetry of the Universe requires the lightest right-handed neutrino mass to lie in the range $0.13\,\text{TeV} \le m_{N_1} \le 230\,\text{TeV}$.
\end{abstract}

\maketitle

\section{Introduction}
\label{sec:Introduction}

Observational cosmology indicates that there is a small but non-zero asymmetry between baryons and anti-baryons in the Universe. Furthermore, this asymmetry is present across our entire Universe, that is, it is overwhelmingly dominated by matter and lacking in, for example, galaxies of antimatter \cite{no_antimatter_galaxies_paper} or other such large scale heterogeneities. This implies that there is some mechanism that allowed for asymmetric production of matter and antimatter in the early universe. This unknown mechanism is then responsible for the net baryon asymmetry observed by Planck of 
\begin{equation}
\eta_B=\dfrac{n_B-n_{\bar{B}}}{n_\gamma}=(6.12\pm0.04)\times10^{-10}
\end{equation}
where $n_B(n_{\bar{B}})$ is the baryon(antibaryon) number density at present time and $ n_\gamma $ is the photon number density\cite{Planck...641A...6P}. According to Sakharov\cite{Textbook_Fukugita:2003neutrino,Sakharov:1967dj}, the baryon asymmetry of the universe may be explained if three conditions are met:

\begin{enumerate}
\item There is a fundamental baryon number violating process,
\item C and CP invariance are violated at the same time,
\item There is a deviation from thermal equilibrium acting on the baryon number violating process.
\end{enumerate}

Fukugita and Yanagida pointed out\cite{Lepto_paper_original_Fukugita:1986hr,Textbook_Fukugita:2003neutrino} that the out-of-equilibrium decay of heavy right-handed (RH) Majorana neutrinos can generate a net lepton asymmetry at relatively low temperatures --- which may then be converted into baryon asymmetry via the $B+L$ violating sphaleron process in such a way that all three of the Sakharov conditions are met. Within leptogenesis models the Sakharov conditions then, by constraining the properties of the heavy RH neutrinos, in turn constrain the low-energy light neutrino properties. 

The simplest extension involving heavy RH Majorana neutrinos is the type-I seesaw model \cite{Minkowski1977,GellMann1979,Yanagida1979,Mohapatra1980,Mohapatra:2007}. While the exchange of heavy RH neutrinos naturally explains the extreme lightness of the observed neutrinos, the model suffers from a vast parameter space, which limits its predictive power.

An attractive idea to overcome this limitation is Sequential Dominance (SD) \cite{King_nice_Vev_search_Master_formula_paper_king2013minimalpredictiveseesawmodel}. SD is a framework for constructing the Dirac mass matrix in which multiple RH neutrinos contribute step-by-step (hierarchically) to generate the light neutrino masses and large mixing angles. Specifically, it imposes a textural structure in which a particular RH neutrino gives a dominant contribution to a specific light neutrino mass. By adopting this structure, the mass hierarchy of neutrinos and the large leptonic mixing angles can be naturally explained without requiring unnatural fine-tuning of parameters.

In this study, we extend phenomenological model construction assuming SD to scenarios with degenerate RH neutrino masses. Unlike standard SD, where the mass hierarchy is driven by the RH neutrino masses, our framework generates the hierarchy solely through Yukawa couplings. Because this setup accommodates resonant leptogenesis as a mechanism for producing the baryon asymmetry of the Universe (BAU), we designate it as Resonant Sequential Dominance (RSD). 

We investigate Dirac mass matrix textures compatible with current neutrino data within RSD, utilizing a novel neutrino mixing matrix ansatz termed TBC3, which is proposed in this work. To identify Yukawa couplings --- arising from the vacuum expectation values (VEVs) of flavons --- that yield viable predictions and take simple values as starting points for fundamental models, we employ King's Master Formula~\cite{King_nice_Vev_search_Master_formula_paper_king2013minimalpredictiveseesawmodel}. This approach allows two of the RH neutrino masses to be nearly degenerate, thereby enabling resonant leptogenesis. Here, ``simple'' refers to small integers ($0, 1, 2, 3, \dots$) and the fractions, roots, and angles constructed from them (e.g., $3/2$, $\sqrt{2}$, $\sqrt{3}$, $\pi/2$). In this sense, $\pi/3$ is simpler than $7\pi/24$, and both are significantly simpler than, for instance, $0.3217$. 

Furthermore, we present an $A_4 \times (Z_3)^4 \times (Z_2)^2$-symmetric Lagrangian that remains consistent with neutrino oscillation data while successfully driving resonant leptogenesis. Finally, we find that explaining the observed baryon asymmetry of the Universe requires the lightest RH neutrino mass to lie within the range $0.13\,\text{TeV} \le m_{N_1} \le 230\,\text{TeV}$.

The rest of this paper is laid out as follows. In Section~2 we give the details of the TBC3 ansatz, its predictions, and investigate how much the second-order expansion deviates from unitarity. In Section~3 we discuss how the important phenomenological features of an SD model may be maintained in the presence of degenerate right-handed neutrino masses by employing RSD. In Section~4 we detail the way we searched for viable simple ratios of the Dirac neutrino mass matrix elements and their unremovable complex phases via King's Master Formula and $\chi^2$ analysis. In Section~5 we present an explicit $A_4\times (Z_3)^4\times(Z_2)^2$-symmetric Lagrangian which admits both RSD and resonant leptogenesis. In Section~6 we establish the connection between RSD parameters and the underlying Lagrangian, as well as discuss the limits of parameter space determined by their connection. We then make numerical predictions for leptogenesis using the best vacuum alignment discovered in Section~4 and demonstrate that it may lead to the correct BAU. In Section~7 we discuss results, present conclusions, and note avenues for potential future research.

\section{PMNS Matrix}
\subsection{Neutrino Oscillations}
The low-energy properties of light left-handed neutrinos are frequently measured by observing their flavor oscillations --- a phenomenon which arises in the lepton sector due to neutrinos being massive, directly analogous to quark mixing described by the Cabibbo-Kobayashi-Masukawa (CKM) matrix\cite{Cabibbo1963,Kobayashi1973}. This leads to a lack of simultaneous diagonalizability of the charged-lepton and neutral-lepton mass matrices. In the diagonal charged-lepton mass matrix basis the flavor eigenstates and mass eigenstates of active light neutrinos in the Standard Model (SM) are related by the unitary Pontecorvo-Maki-Nakagawa-Sakata~\cite{Pontecorvo1958,Maki1962} (PMNS)  mixing matrix U, whose typical parameterization, sans Majorana phases, is given as
\begin{widetext}
\begin{align}
U  
&=\left(
\begin{matrix}
U_{e1} & U_{e2} & U_{e3}  \\
U_{\mu 1} & U_{\mu 2} & U_{\mu 3}  \\
U_{\tau 1} & U_{\tau 2} & U_{\tau 3}  \\
\end{matrix} 
\right)
=
\left ( 
\begin{array}{ccc}
c_{12}c_{13} & s_{12}c_{13} & s_{13} e^{-i\delta^{CP}} \\
- s_{12}c_{23} - c_{12}s_{23}s_{13} e^{i\delta^{CP}} & c_{12}c_{23} - s_{12}s_{23}s_{13}e^{i\delta^{CP}} & s_{23}c_{13} \\
s_{12}s_{23} - c_{12}c_{23}s_{13}e^{i\delta^{CP}} & - c_{12}s_{23} - s_{12}c_{23}s_{13}e^{i\delta^{CP}} & c_{23}c_{13} \\
\end{array}
\right),
\end{align}
\end{widetext}
where in the usual fashion $c_{ij}=\cos\theta_{ij}$, $s_{ij}=\sin\theta_{ij}$, and $\delta^{CP}$ is the observable complex Dirac phase. As per the 2025 NuFIT global best-fit of the current data obtained from neutrino oscillation experiments, including Super-Kamiokande atmospheric neutrino data, in the normal mass ordering (NO) scenario $m_1 < m_2 < m_3$, the mixing angles and the squared mass differences $\Delta m_{ij}^2 = m^2_i - m^2_j$ are\cite{Esteban2024JHEP}:
\begin{align}
s_{12}^2 &= 0.3088^{+0.0067}_{-0.0066} \ (0.2893 \rightarrow 0.3295), \nonumber \quad\\
s_{23}^2 &= 0.470^{+0.017}_{-0.014} \ (0.435 \rightarrow 0.584), \nonumber\\
s_{13}^2 &= 0.02248^{+0.00055}_{-0.00059} \ (0.02064 \rightarrow 0.02418),\nonumber \quad\\
{\delta^{CP}}^\circ &= 212^{+26}_{-36} \ (125 \rightarrow 365),\nonumber \\
\frac{\Delta m_{21}^2}{10^{-5} {\rm \,eV^2}} &= 7.537^{+0.094}_{-0.10} \ (7.236 \rightarrow 7.823),\nonumber \quad\\
\frac{\Delta m_{32}^2}{10^{-3} {\rm \,eV^2}} &= +2.511^{+0.021}_{-0.020} \ (+2.450 \rightarrow +2.576).
\label{Eq:NuFIT_NO}
\end{align}
Here $\pm$ represents the $1 \sigma$ region and the parentheses denote the $3 \sigma$ region. In this work only the NO case is considered --- which continues to be the increasingly favored scenario of the two as evidenced by, for example, the 2026 analysis of DESI DR2 data in Ref.~\cite{NO_Evidence_recent_analysis_jimenez2026evidenceevidentdecisivecosmological} which claims ``decisive cosmological evidence" for NO. For the purposes of completeness the values in the inverted mass ordering (IO) scenario where $m_3 < m_1 \lesssim m_2$ are given: 
\begin{align}
s_{12}^2 &= 0.3088^{+0.0067}_{-0.0066} \ (0.2893 \rightarrow 0.3295),\nonumber \quad\\
s_{23}^2 &= 0.550^{+0.013}_{-0.016} \ (0.439 \rightarrow 0.584), \nonumber\\
s_{13}^2 &= 0.02262^{+0.00057}_{-0.00056} \ (0.02093 \rightarrow 0.02441),\nonumber\quad\\
{\delta^{CP}}^\circ &= 274^{+22}_{-25} \ (203 \rightarrow 335),\nonumber \\
\frac{\Delta m_{21}^2}{10^{-5} {\rm\,eV^2}} &= 7.537^{+0.094}_{-0.10} \ (7.236 \rightarrow 7.822),\nonumber\quad\\
\frac{\Delta m_{32}^2}{10^{-3} {\rm \,eV^2}}& = -2.483^{+0.020}_{-0.020} \ (-2.547 \rightarrow -2.421).
\label{Eq:NuFIT_IO}
\end{align}

 The most recent increase in the precision of $\theta_{12}$ is due to the late 2025 measurement of the solar angle published in the first results of the JUNO collaboration\cite{JUNO_first_results_abusleme2025measurementreactorneutrinooscillations}
\begin{equation}
\sin^2\theta_{12,\:\mathrm{JUNO}}= 0.3092 \pm 0.0087\label{Eq:Juno_theta_12}.
\end{equation}
In Section~\ref{sec:sub:tbc3} the TBC3 matrix ansatz is presented, the goal of which is to match these experimental observations. 

\subsection{TBM and TM$_2$}
\label{sec:PMNS Matrix}

Up until just over a decade ago, oscillation data was in good agreement with the well-known Tri-Bi-Maximal (TBM)~\cite{Harrison:2002er} ansatz for the PMNS matrix 
\begin{eqnarray}
U_{\rm TBM}=\left(
\begin{array}{ccc}
\sqrt{\frac{2}{3}} & \sqrt{\frac{1}{3}}  & 0 \\
-\sqrt{\frac{1}{6}} &  \sqrt{\frac{1}{3}}  &-\sqrt{\frac{1}{2}}  \\
-\sqrt{\frac{1}{6}} &  \sqrt{\frac{1}{3}}  &\sqrt{\frac{1}{2}}
\end{array}
\right).
\label{Eq:UTBM}
\end{eqnarray}
One attractive feature of this ansatz is that such a mixing pattern can arise naturally by enlarging the symmetry of the SM by a family symmetry given by a discrete finite group as small as $S_4$ or its subgroup $A_4$~\cite{Lam_2008_Unique_horizontal_symmetry_of_leptons}. Family (or horizontal) symmetry refers to a symmetry that connects the different families/generations of particles such as (e, $\mu$, $\tau$) --- and since the different generations are typically grouped in columns in diagrammatic representations of the SM they are also referred to as horizontal symmetries. 

In direct model building approaches, such a family symmetry is expected to be respected by the entire Lagrangian at high energies, but below the electroweak (EW) symmetry breaking scale it is broken down to remnant symmetries which the charged- and neutral-lepton sectors respect independently. For example, TBM mixing can arise from an $S_4=\{F,G_1,G_2,G_3\}$-symmetric model which breaks such that the charged-lepton sector is $F$-symmetric and the neutral-lepton sector is symmetric under all remaining generators $G_i$, in which case the columns, $U_{\mathrm{TBM},i}$, are the invariant eigenvectors of $G_i$ with eigenvalue $+1$~\cite{CS_Lam_Group_theory_and_dynamics_of_neutrino_mixing}. 

The mixing angles predicted by such an approach were initially in agreement with experimental observations, but from 2011-2012 the T2K~\cite{Abe:2011sj}, MINOS \cite{MINOS:2011amj}, Double Chooz\cite{DOUBLE_CHOOZ_Abe2012}, Daya Bay\cite{DAYA_BAY_An2012}, and RENO\cite{RENO_Kim2012} experiments demonstrated that the reactor angle, identified with the $U_{e3}$ element of the PMNS matrix, was non-zero, with the current best-fit value being $s_{13}\approx8.5^\circ$. For some discussion of model building approaches in light of the non-zero reactor angle see Ref.~\cite{PhysRevD.85.031903,King2011TrimaximalNM,King_Review}. One common method is, for example, to construct an $S_4$-symmetric model whose remnant symmetry in the neutral-lepton sector is only one of the three generators $G_i$~\cite{CS_Lam_Group_theory_and_dynamics_of_neutrino_mixing}. This leads to one column being fixed and the other two columns containing unfixed parameters subject only to unitarity constraints. Depending on which $G_i$ is chosen, these are referred to as TM$_i$\cite{TMi_Albright_2009}. For example, resonant leptogenesis using TM$_1$ produced from an $S_4$ model was studied in Ref.~\cite{Thapa_2021}. Choosing $G_2$ may naturally yield TM$_2$ which can also accommodate the reactor angle
\begin{equation}
U_{\rm TM_2}=\left(
\begin{matrix}
\sqrt{\frac{2}{3}}\cos\theta &  \sqrt{\frac{1}{3}} & \sqrt{\frac{2}{3}}\sin\theta  \\
-\frac{\cos\theta}{\sqrt{6}} + \frac{e^{-i\phi}\sin\theta}{\sqrt{2}}  &\sqrt{\frac{1}{3}}  & -\frac{\sin\theta}{\sqrt{6}} - \frac{e^{-i\phi}\cos\theta}{\sqrt{2}} \\
-\frac{\cos\theta}{\sqrt{6}} - \frac{e^{-i\phi}\sin\theta}{\sqrt{2}}  & \sqrt{\frac{1}{3}}  & -\frac{\sin\theta}{\sqrt{6}}  + \frac{e^{-i\phi}\cos\theta}{\sqrt{2}} \\
\end{matrix}
\right)
\label{Eq:UTM2},
\end{equation}
where $\theta$ and $\phi$ are model parameters. This mixing matrix contains the following strict constraint
\begin{align}
s^2_{12} &= \frac{1}{3(1-s^2_{13})}
\end{align}
which sets the limit $s^2_{12}\geq1/3$. TM$_2$ and non-resonant leptogenesis has been studied in, for example, Ref.~\cite{TM2_nonresonant_He_2007}.

Up until very recently this was still in good agreement with measurements of the solar angle, but now, to high precision, this is in strong conflict with the 2025 first results from JUNO given in Eq.~(\ref{Eq:Juno_theta_12}). In combination with the non-zero reactor angle, a solar angle below $1/3$ further complicates the situation and demands additional modification to the TBM/TM$_2$ approaches, or some new starting point entirely. Previously, the authors of this work proposed one such modification to TM$_2$~\cite{Fodroci_our_paper_10.1093/ptep/ptag044} wherein the structure of the PMNS matrix is modified directly in order to allow fitting with experimental data. Other recent approaches include tactics such as corrections stemming from a non-diagonal charged-lepton mass matrix as in Ref.~\cite{Cites_Our_paper_singh2026reconcilingtm2mixinglma}. 

In the context of indirect vs direct model building the situation is summarized as follows. For a direct model, where the high-energy Lagrangian's symmetry breaks to remnant symmetries which are independently respected in the charged- and neutral-lepton sectors, the typical way now to achieve agreement with the current experimental data is to use very large groups. For example, the group $\Delta(2904)$ can give a good fit to the reactor angle~\cite{KING_Delta_6_n_sq_BIG_GROUP_King:2013iva,King_Review}. A systematic scan of all groups up to order 1536 performed in Ref.~\cite {Holthausen:2012wt} yielded only 3 groups compatible with the $3\sigma$ range of 2013 mixing best-fits.

In the case of indirect models, where the remnant symmetry in the neutral-lepton sector is entirely broken, then it usually is a matter of choosing and justifying special Vacuum Expectation Values (VEVs) of flavons present in higher-dimensional operators responsible for the lepton masses --- it is this latter approach which is focused on in this work. 

\subsection{TBC3 Ansatz}\label{sec:sub:tbc3}
 In light of the JUNO result and the global best-fit values in Eqs.(\ref{Eq:NuFIT_NO}) and/or (\ref{Eq:NuFIT_IO}), we suggest a new ansatz for producing a PMNS matrix which matches experimental data. We start with S. F. King's\cite{King_PMNS_Matrix_expansion_origin_paper} parameterization of $U_{\mathrm{TBM}}$ in terms of deviation from the TBM solar, atmospheric, and reactor angles --- given by $s$, $a$, and $r$, respectively. In his parameterization the mixing angles are given by 
\begin{align}
 s_{12} = \frac{1}{\sqrt{3}}(1+s),\:\:\:\:s_{23} = \frac{1}{\sqrt{2}}(1+a),\:\:\:\:s_{13} = \frac{r}{\sqrt{2}},
 \label{a,s,r,deviations}
\end{align}
and to the first-order, neglecting Majorana phases, the mixing matrix is 
\begin{widetext}
\begin{eqnarray}
U \approx
\left( \begin{array}{ccc}
\sqrt{\frac{2}{3}}(1-\frac{1}{2}s)  & \frac{1}{\sqrt{3}}(1+s) & \frac{1}{\sqrt{2}}re^{-i\delta } \\
-\frac{1}{\sqrt{6}}(1+s-a + re^{i\delta })  & \frac{1}{\sqrt{3}}(1-\frac{1}{2}s-a- \frac{1}{2}re^{i\delta })
& \frac{1}{\sqrt{2}}(1+a) \\
\frac{1}{\sqrt{6}}(1+s+a- re^{i\delta })  & -\frac{1}{\sqrt{3}}(1-\frac{1}{2}s+a+ \frac{1}{2}re^{i\delta })
 & \frac{1}{\sqrt{2}}(1-a)
\end{array}
\right).
\label{PMNS_1st_order}
\end{eqnarray}
\end{widetext}
The second-order corrections derived in Ref.~\cite{King_PMNS_Matrix_expansion_origin_paper} are
\begin{eqnarray}
\Delta U_{e1} & \approx & \sqrt{\frac{2}{3}}\left(-\frac{1}{4}r^2 -\frac{3}{8}s^2\right), \nonumber \\
\Delta U_{e2} & \approx & \frac{1}{\sqrt{3}}\left( -\frac{1}{4}r^2\right), \nonumber \\
\Delta U_{e3} & \approx & 0,\nonumber \\
\Delta U_{\mu 1} & \approx & -\frac{1}{\sqrt{6}}\left(\frac{1}{2}rse^{i\delta }  -rae^{i\delta }  +sa +a^2\right), \nonumber \\
\Delta U_{\mu 2} & \approx & \frac{1}{\sqrt{3}}\left(-\frac{1}{2}rse^{i\delta }  -\frac{1}{2}rae^{i\delta }  +\frac{1}{2}sa - \frac{3}{8}s^2-a^2\right),
\nonumber \\
\Delta U_{\mu 3} & \approx & \frac{1}{\sqrt{2}}\left(-\frac{1}{4}r^2\right),
\nonumber \\
\Delta U_{\tau 1} & \approx &
\frac{1}{\sqrt{6}}\left(\frac{1}{2}rse^{i\delta }  +rae^{i\delta }  +sa\right),
\nonumber \\
\Delta U_{\tau 2} & \approx &
-\frac{1}{\sqrt{3}}\left(\frac{1}{2}rse^{i\delta }  -\frac{1}{2}rae^{i\delta }  -\frac{1}{2}sa - \frac{3}{8}s^2\right),
\nonumber \\
\Delta U_{\tau 3} & \approx &
\frac{1}{\sqrt{2}}\left(-\frac{1}{4}r^2 -a^2\right).
\label{PMNS_2nd_order_PMNS_King_Corrections}
\end{eqnarray}
This type of approach is frequently found in the literature and is inspired by the remarkably successful Wolfenstein parameterization~\cite{Wolfenstein1983} of the CKM matrix. In said parameterization, the CKM matrix is expanded in terms of the small Cabibbo parameter~\cite{Cabibbo1963} such that the zeroth order term is the identity matrix. King later proposed the above analogous expansion for the PMNS matrix which takes the zeroth order term to be the TBM mixing matrix $U_{\rm TBM}$.

Taking $\lambda$ to represent the Cabibbo parameter, we adopt the following ansatz, which, following Ref.~\cite{King_nice_Vev_search_Master_formula_paper_king2013minimalpredictiveseesawmodel}, we denote TBC3: 
\begin{equation}
s=a=-\dfrac{1}{3\sqrt{3}}\:\lambda,\:\:\:\:r=\dfrac{2}{3}\sqrt{2}\:\lambda.
\label{TBC3_ansatz}
\end{equation}
This is motivated by the comparable size of the solar and atmospheric angles' deviations from their TBM values and the desire to reduce the number of model parameters. Furthermore, these values can achieve agreement with NuFIT $1\sigma$ allowed regions for the mixing angles while only employing the products of simple ratios with the Cabibbo parameter. Via Eq.~(\ref{a,s,r,deviations}) this leads to the previously stated first-order values of the mixing angles
\begin{equation}
s_{12}=\dfrac{1}{\sqrt{3}}\left(1-\dfrac{\lambda}{3\sqrt3}\right),\:\:\:\:s_{23}=\dfrac{1}{\sqrt{2}}\left(1-\dfrac{\lambda}{3\sqrt3}\right),\:\:\:\:
s_{13}=\dfrac{2}{3}\lambda.
\label{angles_in_terms_of_TBC3_ansatz}
\end{equation}
Also to the first order,
\begin{widetext}
\begin{eqnarray} U_{\rm TBC3} \approx
\left( \begin{array}{ccc}
\dfrac{\sqrt{2}}{\sqrt{3}}\:\biggr(\:1+\dfrac{1}{6\sqrt{3}}\:\lambda\:\biggr)  & \dfrac{1}{\sqrt{3}}\:\biggr(\:1-\dfrac{1}{3\sqrt{3}}\:\lambda\:\biggr) & \dfrac{2}{3}\lambda\: e^{-i\delta} \\
-\dfrac{1}{\sqrt{6}}\:\biggr(\:1+\dfrac{2}{3}\sqrt{2}\:\lambda\: e^{i\delta}\:\biggr) & \dfrac{1}{\sqrt{3}}\:\biggr(\:1+\dfrac{1}{2\sqrt{3}}\:\lambda-\dfrac{1}{3}\sqrt{2}\:\lambda\: e^{i\delta}\:\biggr) & \dfrac{1}{\sqrt{2}}\:\biggr(\:1-\dfrac{1}{3\sqrt{3}}\:\lambda\:\biggr) \\
\dfrac{1}{\sqrt{6}}\:\biggr(\:1-\dfrac{2}{3\sqrt{3}}\:\lambda-\dfrac{2}{3}\sqrt{2}\:\lambda \:e^{i\delta}\:\biggr)   & -\dfrac{1}{\sqrt{3}}\:\biggr(\:1-\dfrac{1}{6\sqrt{3}}\lambda+\dfrac{1}{3}\sqrt{2}\:\lambda\: e^{i \delta}\:\biggr)  & \dfrac{1}{\sqrt{2}}\biggr(\:1+\dfrac{1}{3\sqrt{3}}\:\lambda\:\biggr)\\
\end{array}
\right).
\label{TBC3_order_1}
\end{eqnarray}
\end{widetext}

We note, and will demonstrate later, that the equivalence of the atmospheric and solar deviations, $s=a$, comes at the cost of reducing the ability to find VEVs of flavons that will produce $1\sigma$ agreement simultaneously with all mixing angles, the Dirac phase, and mass-squared differences --- but that disagreement is sufficiently minimal that this interesting ansatz is maintained throughout this work. 

Taking the PDG value of $\lambda\approx0.22501$~\cite{PDG_PhysRevD.110.030001} Eq.~(\ref{angles_in_terms_of_TBC3_ansatz}) gives us 
\begin{equation}
\theta_{12}=33.53^\circ\:\:\:\:\theta_{23}=42.57^\circ,\:\:\:\:\theta_{13}=8.63^\circ,
\label{TBC_raw_angle_predictions}
\end{equation}
which is in good agreement with experimental data. Furthermore, after including the second-order corrections in Eq.~(\ref{PMNS_2nd_order_PMNS_King_Corrections}), this yields the following magnitudes of each PMNS matrix element for $\delta^{CP}=-\pi$:
\begin{eqnarray}
\bigg|U_{\rm TBC3}(\delta^{CP}=-\pi)\bigg|\approx
\left( \begin{array}{ccc}
0.824  & 0.546 & 0.150 \\
0.321  & 0.670 & 0.668 \\
0.466  & 0.503 & 0.728 \\
\end{array}
\right).
\label{abs of TBC3}
\end{eqnarray}
This agrees well with the current best $3\sigma$ NuFIT values with SK atmospheric data\cite{Esteban2024JHEP} for the magnitude of the elements of the PMNS matrix
\begin{widetext}
\begin{equation}
|U|_{3\sigma}^{\text{IC24 with SK-atm}} = \begin{pmatrix} 
\:\:\:0.810 \rightarrow 0.834 & 0.532 \rightarrow 0.568 & 0.144 \rightarrow 0.156\:\:\: \\
\:\:\:0.259 \rightarrow 0.498 & 0.514 \rightarrow 0.678 & 0.652 \rightarrow 0.756\:\:\: \\
\:\:\:0.280 \rightarrow 0.512 & 0.488 \rightarrow 0.659 & 0.637 \rightarrow 0.743\:\:\: 
\end{pmatrix}.
\end{equation}
\end{widetext}

\subsection{Unitarity of TBC3}
\begin{figure}[htbp]
    \centering
    \includegraphics[width=1\linewidth]{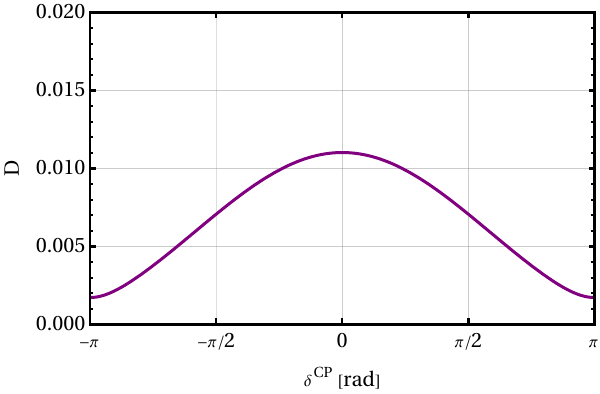}
    \caption{ Deviation from unitarity of TBC3 ansatz in terms of the Frobenius Norm as a function of the CP-violating phase $\delta^{CP}$.}
    \label{fig:Unitarity}
\end{figure}

Given that we are using an expansion about the small parameters $a$, $s$, and $r$, which is truncated at the second-order to compute the PMNS matrix elements, it is expected that there should be some small deviation from exact unitarity. To parameterize deviation from unitarity we utilize the Frobenius norm, which for a general $m\times n$ matrix $A$ is given by  
\begin{equation}
    \lVert A\rVert_F=\sqrt{\sum^m_i \sum^n_j |a_{ij}|^2}.
\end{equation}
Using this, we define the following parameter as a measure of deviation of $U_{\rm TBC3}$ from unitarity
\begin{equation}
    D(\delta^{CP})\equiv\lVert I_3- U_{\rm TBC3}U_{\rm TBC3}^\dagger\rVert_F.
\end{equation} 
This is plotted for the TBC3 ansatz as a function of $\delta^{CP}$ in Figure~\ref{fig:Unitarity}. These sub-$1\%$ deviations from unitarity are acceptable given that they are well below current experimental precision. Furthermore the deviations are smallest in the neighborhood of the best-fit of $\delta^{CP}$.

\section{Resonant Sequential Dominance}
\label{sec:Resonant Sequential Dominance}
\subsection{Conventional SD}
One of the central motivations for extending the SM is explaining the observed smallness of neutrino masses. One such popular extension is the well-known seesaw mechanism\cite{Minkowski1977,GellMann1979,Yanagida1979,Mohapatra1980,Valle_Mohapatra:1986bd,Valle_Schechter:1980gr,Valle_Schechter:1981cv} wherein the smallness of the left-handed active neutrinos is naturally explained via the introduction of some number of very heavy fields which, after being integrated out, naturally suppress the left-handed neutrino masses. In particular, we focus on the Type-I seesaw model~\cite{Minkowski1977,GellMann1979,Yanagida1979,Mohapatra1980,Mohapatra:2007} where the heavy fields are the RH Majorana neutrinos. Their Majorana nature allows them to be the only fundamental particle that can form a singlet mass term under the SM symmetries and, after integrating these heavy neutrinos out at low energies, their heavy masses naturally suppress the effective light neutrino mass terms through the seesaw mechanism. This removes the need for putting in unnaturally small Yukawa couplings by hand. In the basis where the charged-lepton mass matrix, $M_{\rm E}$, and Majorana neutrino mass matrix, $M_{\rm R}$, are both diagonal, a typical lepton mass Lagrangian term after EW symmetry breaking can be written as
\begin{equation}
{\cal L}=-  \overline{l}_{l} M_{\rm E} l_{R}  -  \overline{\nu}_{ L} m_{\rm D} N_{R} 
- \frac{1}{2}\overline{N}^{ c}_{ R} M_{\rm R} N_{ R} 
+ {\text h.c.}\; ,
\end{equation}
where $l_L=(e_L,\mu_L,\tau_L)$, $l_R=(e_R,\mu_R,\tau_R)$, $\nu_L=(\nu_e,\nu_\mu,\nu_\tau)$, and $N_R$ comprises the, in this study, three RH neutrinos $N_R=(N_{1},N_{2}, N_{3})$~\cite{King_nice_Vev_search_Master_formula_paper_king2013minimalpredictiveseesawmodel}. Naturally, $l_l$ and $\nu_l$ form the SU(2) doublet $L=(\nu_L,l_l)^T$ whereas all of the RH fields are SM singlets. $m_{\rm D}$ denotes the Dirac neutrino mass matrix.  Explicitly, the two diagonal mass matrices are
\begin{equation}
M_{\rm E}=
\left( \begin{array}{ccc}
m_{ e} & 0 & 0    \\
0 & m_{ \mu} & 0 \\
0 & 0 & m_{ \tau}
\end{array}
\right), \
M_{\rm R}=
\left( \begin{array}{ccc}
M_{1} & 0 & 0    \\
0 & M_{2} & 0 \\
0 & 0 & [M_{3}]
\end{array}
\right).
\label{seq1}
\end{equation}
where all six eigenvalues are, without loss of generality, taken to be positive and real. The brackets here around $M_3$, in keeping with Ref.~\cite{King_nice_Vev_search_Master_formula_paper_king2013minimalpredictiveseesawmodel}, indicate that the mass of RH neutrino $N_3$ is ultra-heavy, with a mass at the typical scale of a Grand Unified Theory ($M_3\simeq10^{15}$\,GeV). 

This allows $N_3$ to be integrated out well above the leptogenesis scale and as such it has negligible effects on low-energy observables. Via the Type-I seesaw mechanism, the light neutrino mass matrix is 
\begin{equation}
    m_{\nu}=m_{\rm D}M_{\rm R}^{-1}{m_{\rm D}}^T.
    \label{see_saw_mechanism_positive_right_habd_side}
\end{equation}
The convention where the right-hand side is positive is chosen so that certain parameters of interest later come out explicitly real. Additionally, the Dirac neutrino mass matrix is
\begin{align}
m_{\rm D}&=
\left( \begin{array}{ccc}
m_{\rm{D},\it e \rm, atm} & m_{\rm D, \it e \rm, sol} & [m_{\rm D, \it e \rm, dec}]   \\
m_{\rm D, \it \mu \rm, atm} & m_{\rm D, \it \mu \rm, sol} & [m_{\rm D, \it \mu \rm, dec}] \\
m_{\rm D, \it \tau \rm, atm} & m_{\rm D, \it \tau \rm, sol} & [m_{\rm D, \it \tau \rm, dec}]
\end{array}
\right)\notag\\
&\equiv \left( \begin{array}{ccc}
m_{\rm D, atm} & m_{\rm D, sol} & [m_{\rm D, dec}]  
\end{array}
\right).
\label{dirac}
\end{align}
The brackets again indicate that these elements may be neglected due to the extreme mass of $N_3$. The column vectors $m_{\rm D,atm}$, $m_{\rm D,sol}$, $m_{\rm D,dec}$ are defined in such a way that one may identify the terms containing $m_{\rm atm}$ as those primarily responsible for the $m_3$ mass, terms containing $m_{\rm sol}$ as primarily responsible for $m_2$, and the almost decoupled $m_{\rm dec}$ as that primarily responsible for $m_1\approx0$~\cite{King_nice_Vev_search_Master_formula_paper_king2013minimalpredictiveseesawmodel}. This leads to the effective light neutrino mass Lagrangian term
\begin{align}
\mathcal{L}^\nu_{eff} =
 &\frac{(\overline{\nu}_L m_{\rm D, atm})(m_{\rm D, atm}^{T} \nu_L^c)}{M_{\rm atm}}
 +\frac{(\overline{\nu}_L m_{\rm D, sol})(m_{\rm D, sol}^{T} \nu_L^c)}{M_{\rm sol}}\notag\\
\ &+\left[ \ \frac{(\overline{\nu}_L m_{\rm D, dec})(m_{\rm D, dec}^{T} \nu_L^c)}{M_{\rm dec}} \ \right] \label{leff}.
\end{align}
In the case that
\begin{equation}
\frac{m_{\rm D, atm}m_{\rm D, atm}^T}{M_{\rm atm}} \gg
\frac{m_{\rm D, sol}m_{\rm D, sol}^T}{M_{\rm sol}} \ \left[ \ \gg
\frac{m_{\rm D, dec}m_{\rm D, dec}^T}{M_{\rm dec}}\ \right] \, ,
\label{SD1}
\end{equation}
this yields the normally ordered neutrino mass hierarchy
\begin{equation}
\label{normal2}
m_3 \gg  m_2 \ \left[\  \gg m_1  \ \right],
\end{equation}
and is referred to as SD~\cite{Bjorkeroth2014CSDn,King2005CSD,King_nice_Vev_search_Master_formula_paper_king2013minimalpredictiveseesawmodel}. This mechanism provides an elegant means of explaining both the smallness and the ordering of the neutrino masses in a way where each of the columns of the Dirac neutrino mass matrix primarily drives one of the three mass-squared differences. While it does not place any constraints on the ordering of the Majorana masses, it does require that, under some ordering, the $M_{1},M_{2}, M_{3}$ obey 
\begin{equation}
    M_{\rm atm}<M_{\rm sol}<M_{\rm dec}. 
\end{equation}
That is, it is not necessarily required that, say, $M_{1} = M_{\rm atm}$.

\subsection{Resonant SD}
In this case, we are interested in developing a model which may lead to resonant leptogenesis. This requires that there are at least two nearly degenerate RH neutrino masses that differ at high energies only by a small parameter $\Delta$ proportional to half the decay width of the heavy RH neutrinos $\Gamma_{N_i}\:/2$. This small splitting leads to a resonant enhancement of loop-level CP-violating decays of the heavy RH neutrinos that allows the generation of sufficient lepton asymmetry in the early universe with RH neutrino masses as low as $1$ TeV~\cite{Resonant_lepto_Pilaftsis_2004}. Therefore, with $M_1=M_{\rm atm}$, $M_2=M_{\rm sol}$, $M_3=M_{\rm dec}$, we choose the ordering
\begin{equation}
    M_1\approx M_2\ll M_3. 
\end{equation}

This ordering is not compatible with the typical realization of SD via the heavy RH neutrino mass hierarchy. In order to rectify this we assume that, after EW symmetry breaking and spontaneous flavor symmetry breaking, Dirac masses arise from higher-dimensional operators of the form 
\begin{equation}
   m_{\mathrm{D},ij}=\dfrac{v}{\sqrt{2}}\dfrac{y_{\mathrm{D}_i}}{\Lambda}\braket{\phi_{ij}}=\dfrac{v}{\sqrt{2}}Y_{ij}.
\end{equation}
The necessary features of SD in Eq.~(\ref{SD1}) are then achieved via only the Yukawa couplings by assuming $y_{\mathrm{D}_1}<y_{\mathrm{D}_2}$ --- with no particular constraints on $y_{\mathrm{D}_3}$ due to the largeness of $M_3$. In particular, this still allows $m_3 \gg  m_2 \ \left[\  \gg m_1  \ \right]$ and for the geometric structure of the flavon VEVs $\braket{\phi_i}$ to be the primary feature determining the low-energy observables. In this work we do not propose any particular mechanism behind the Yukawa hierarchy, $y_{\mathrm{D}_1}<y_{\mathrm{D}_2}$, but note that, similar to the charged-lepton sector, such an ordering may in principle be obtained through the Froggatt-Nielsen (FN) mechanism~\cite{FROGGATT1979277} or an FN-like approach. 

\begin{figure*}[htbp]
\centering
\begin{tikzpicture}[
    node distance=1.8cm,
    every node/.style={align=center},
    box/.style={
        draw,
        rounded corners,
        minimum width=0.38\textwidth,
        minimum height=1.25cm,
        text width=0.34\textwidth,
        inner sep=6pt
    },
    arrow/.style={
        thick,
        ->,
        >=latex
    }
]
\node at (-5.2,0) {\Large\underline{Conventional SD}};
\node at ( 5.2,0) {\Large\underline{Resonant SD}};
\node[box] (L1) at (-5.2,-1.8)
{
Hierarchical heavy RH neutrino masses
\[
M_{\rm atm}\ll M_{\rm sol}\ll M_{\rm dec}.
\]
};

\node[box] (L2) at (-5.2,-4.4)
{
Light neutrino mass ordering primarily from RH neutrino mass hierarchy
};

\draw[arrow] (L1) -- (L2);
\node[box] (R1) at (5.2,-1.8)
{
At least two RH neutrinos nearly degenerate
\[
M_1\approx M_2\ll M_3.
\]
};

\node[box] (R2) at (5.2,-4.4)
{
Light neutrino mass ordering primarily from Yukawa coupling hierarchy
};

\node[box] (R3) at (5.2,-7.0)
{
Sequential Dominance which allows for resonant leptogenesis
};

\draw[arrow] (R1) -- (R2);
\draw[arrow] (R2) -- (R3);
\end{tikzpicture}

\caption{Comparison between conventional Sequential Dominance (SD) and the Resonant Sequential Dominance (RSD) framework proposed in this work. The light neutrino mass hierarchy in SD arises primarily through the heavy RH neutrino mass hierarchy, whereas in RSD at least two of the RH neutrinos are near degenerate and instead the light neutrino mass hierarchy originates primarily in the Yukawa couplings. Crucially, this permits resonant leptogenesis.}\label{RSD_vs_SD_flowchart}
\end{figure*}
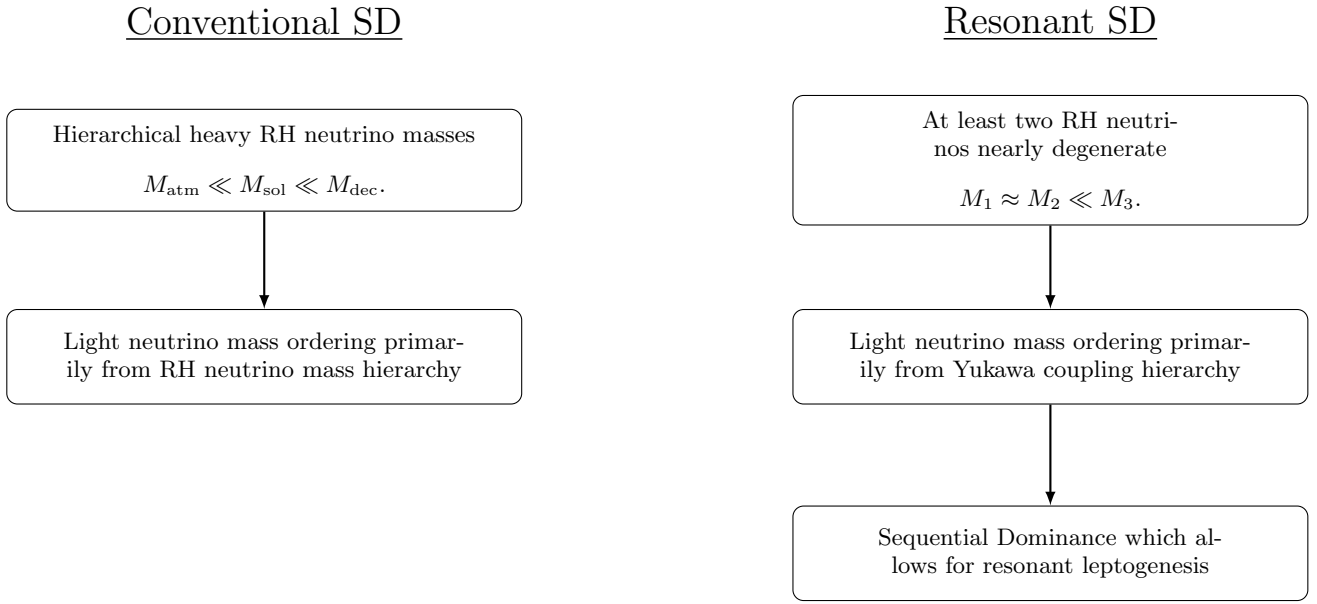

We refer to this realization of SD, in which the hierarchy arises solely through the Dirac Yukawas and not the RH neutrino mass hierarchy, as RSD. This preserves the relationship between the columns of the Dirac neutrino mass matrix and the low-energy observables --- most importantly that each column is primarily responsible for generating one of the mass-squared differences. The conceptual differences between SD and RSD are outlined in Figure \ref{RSD_vs_SD_flowchart}.

\section{Searching for Simple Dirac Neutrino Mass Matrix Elements/Flavon VEVs}
\label{sec:VEV Search}
\subsection{Master Formula}
We work in the basis where the charged-lepton mass matrix is diagonal and assume a texture zero in the first entry of the column corresponding to the lightest right-handed neutrino. We further assume degenerate RH Majorana masses, $M = M_1 = M_2$, so that
\begin{align}
&m_{\rm D}=\left( \begin{array}{cc}
0 & a    \\
e & b  \\
f & c 
\end{array}
\right)=
\left( \begin{array}{cc}
m_{\rm D,\it e \rm, atm} & m_{\rm D, \it e \rm, sol}    \\
m_{\rm D, \it \mu \rm, atm} & m_{\rm D, \it \mu \rm, sol}  \\
m_{\rm D, \it \tau \rm, atm} & m_{\rm D, \it \tau \rm, sol} 
\end{array}
\right),\notag\\ &M_{\rm R}=\left( \begin{array}{cc}
M & 0    \\
0 & M  \\ 
\end{array}\right).
\end{align}
This texture zero is shown in Ref.~\cite{King_nice_Vev_search_Master_formula_paper_king2013minimalpredictiveseesawmodel} to be closely connected to a non-zero reactor angle. We then apply the Master Formula from Ref.~\cite{King_nice_Vev_search_Master_formula_paper_king2013minimalpredictiveseesawmodel} for relating the Dirac mass matrix elements and their complex phases, the Majorana mass matrix, and the elements of the PMNS matrix, $U$,
\begin{align}
m_\nu&=\left( \begin{array}{ccc}
\tilde{a}^2 & \tilde{a}\tilde{b}   &  \tilde{a}\tilde{c}   \\
 \tilde{a}\tilde{b}  &  \tilde{e}^2+\tilde{b}^2  &  \tilde{e}\tilde{f}+ \tilde{b}\tilde{c}  \\
 \tilde{a}\tilde{c} & \tilde{e}\tilde{f}+ \tilde{b}\tilde{c} & \tilde{f}^2+\tilde{c}^2
\end{array}
\right)_{\alpha \beta}\notag\\ 
&=e^{i\beta}m_2U_{\alpha 2}U_{\beta 2}+
m_3U_{\alpha 3}U_{\beta 3},
\label{Master_formula}
\end{align}
where the left-hand-side quantities are defined as
\begin{align}
 &\tilde{a} \equiv \frac{ie^{i\phi_{e}}a}{\sqrt{M}},  \ \  \tilde{b} \equiv \frac{ie^{i\phi_{\mu}}b}{\sqrt{M}},  
 \ \ \tilde{c} \equiv \frac{ie^{i\phi_{\tau}}c}{\sqrt{M}},\notag\\
 &\  \  \tilde{e} \equiv \frac{ie^{i\phi_{\mu}}e}{\sqrt{M}}, \ \  \tilde{f} \equiv \frac{ie^{i\phi_{\tau}}f}{\sqrt{M}}.
 \label{dirac_matrix_elements_with_complex_phase_divided_by_RH_masses}
\end{align}

This Master Formula provides a direct relationship between the low-energy observables and the Dirac mass matrix elements from which they arise and so is an excellent tool to search for simple values of the Yukawa couplings that can reproduce experimental observables. In Ref.~\cite{King_nice_Vev_search_Master_formula_paper_king2013minimalpredictiveseesawmodel} the terms in Eq.~(\ref{dirac_matrix_elements_with_complex_phase_divided_by_RH_masses}) pertaining to the sub-dominant light neutrino mass (those containing $a$, $b$, and $c$) have a different RH neutrino mass in the denominator as opposed to those pertaining to the dominant light neutrino mass (containing $e$ and $f$). It is assumed that, below the leptogenesis scale, the RH neutrino masses are degenerate. Under this change we will show that the predictive power of this Master Formula is maintained, though the interpretation of several model parameters changes (see Eq.~(\ref{light_neutrino_mass_matrix_as_function_of_A,B}) and following discussion).

Having presented the Master Formula we can now state the definitions of the following quantities
\begin{equation}
\begin{array}{ccccl}
Z_1 &\equiv&\dfrac{\tilde{e}}{\tilde{f}}&=&\pm\dfrac{U_{23}U_{12}-U_{13}U_{22}}{U_{33}U_{12}-U_{13}U_{32}},\\
Z_2 &\equiv&\dfrac{\tilde{b}}{\tilde{a}} &=&\dfrac{e^{i\beta}m_2U_{1 2}U_{2 2}+ m_3U_{1 3}U_{2 3}}
{e^{i\beta}m_2U_{1 2}^2+ m_3U_{1 3}^2},\\
Z_3 &\equiv&\dfrac{\tilde{c}}{\tilde{a}}&=&\dfrac{e^{i\beta}m_2U_{1 2}U_{3 2}+ m_3U_{1 3}U_{3 3}}
{e^{i\beta}m_2U_{1 2}^2+ m_3U_{1 3}^2},
\end{array}
\label{z3_explicit_expressions}
\end{equation}
and their moduli in which we are primarily interested
\begin{equation}
|Z_1|=\dfrac{|\tilde{e}|}{|\tilde{f}|},\:\:\:\:|Z_2|=\dfrac{|\tilde{b}|}{|\tilde{a}|},\:\:\:\:|Z_3|=\dfrac{|\tilde{c}|}{|\tilde{a}|}.
\label{Z_i, moduli}
\end{equation}
Additionally, writing the Dirac neutrino mass matrix in terms of the phases
\begin{equation}
\label{basis}
m_{\rm D}=\left( \begin{array}{cc}
0 & |a|e^{i\phi_a}   \\
|e|e^{i\phi_e} & |b|e^{i\phi_b} \\
|f|e^{i\phi_f} & |c|e^{i\phi_c}  
\end{array}
\right),
\end{equation}
we note that only two physical (unremovable) complex phases exist which may be written as 
\begin{equation}
\eta_2=\phi_b-\phi_e=\arg{\tilde{b}}-\arg{\tilde{e}}, 
\ \ \ \ \eta_3= \phi_c-\phi_f=\arg{\tilde{c}}-\arg{\tilde{f}}.
\label{eta_2_eta_3_defintions}
\end{equation}
In all cases, we choose 
\begin{equation}
    \eta_1=\phi_a=0.
\end{equation}
In keeping with Ref.~\cite{King_nice_Vev_search_Master_formula_paper_king2013minimalpredictiveseesawmodel} we compute these phases by first fixing 
\begin{equation}
\arg{\tilde{b}}=\frac{1}{2}\arg{\tilde{b}^2}, 
\ \ \ \ \arg{\tilde{e}}=\frac{1}{2}\arg{\tilde{e}^2}.
\end{equation}
This is then used together with 
\begin{equation}
\arg{\tilde{c}} = \frac{1}{2}\arg{\tilde{c}^2} - \arg\left( \frac{\tilde{a}\tilde{b}}{\tilde{a}\tilde{c}} \right),
\end{equation}
\begin{align}
\arg{\tilde{f}} =&-\arg{\tilde{e}}-\arg{\tilde{b}}-\arg{\tilde{c}}\notag\\&+\arg\frac{1}{2}\left[  
({\tilde{e}}{\tilde{f}}+{\tilde{b}}{\tilde{c}})^2 - {\tilde{e}}^2{\tilde{f}}^2 -  {\tilde{b}}^2{\tilde{c}}^2
\right].
\end{align}
 Also useful for computing $Z_i$, $\eta_i$ are the following identities derived from the Master Formula:
\begin{equation}\begin{array}{lcl}\tilde{a}^2 & =& e^{i\beta}m_2U_{1 2}^2+ m_3U_{1 3}^2,\\
\tilde{a}\tilde{b} & =& e^{i\beta}m_2U_{1 2}U_{2 2}+ m_3U_{1 3}U_{2 3},\\
\tilde{a}\tilde{c} & =& e^{i\beta}m_2U_{1 2}U_{3 2}+ m_3U_{1 3}U_{3 3},\\
\tilde{e}\tilde{f}+\tilde{b}\tilde{c}  & =& e^{i\beta}m_2U_{2 2}U_{3 2}+ m_3U_{2 3}U_{3 3},\\
\tilde{e}^2 + \tilde{b}^2 & =& e^{i\beta}m_2U_{2 2}^2+ m_3U_{2 3}^2,\\
\tilde{f}^2 + \tilde{c}^2 & =& e^{i\beta}m_2U_{3 2}^2+ m_3U_{3 3}^2,\\
\tilde{b}^2 & =&  \dfrac{(\tilde{a}\tilde{b})^2}{\tilde{a}^2},\\
\tilde{c}^2 & =&  \dfrac{(\tilde{a}\tilde{c})^2}{\tilde{a}^2},\\
\tilde{e}^2 & =&  (\tilde{e}^2 + \tilde{b}^2)-\tilde{b}^2,\\
\tilde{f}^2 & =&  (\tilde{f}^2 + \tilde{c}^2)-\tilde{c}^2.
\end{array}\end{equation}
Lastly, for the purposes of numerical calculations, the same approximation is taken as in Ref.~\cite{King_nice_Vev_search_Master_formula_paper_king2013minimalpredictiveseesawmodel} for the mass ratio 
\begin{equation}
\frac{m_2}{m_3}\approx  \frac{\lambda}{ \sqrt{2}} + \frac{1}{3}\lambda^2\approx 0.176,
\label{m2_m3_in_terms_of_cabibo}
\end{equation}
which is still in agreement with both current NuFIT best-fit data on the mass squared differences and the current best-fit value of the Cabibbo parameter.

\subsection{TBC3 Master Formula Calculations and $\chi^2$ Analysis}
We first compute $|Z_1|$ as a function of $\delta^{CP}$ which determines the ratio $e/f$. Then, for a chosen $\delta^{CP}$ we determine $|Z_2|$ and $|Z_3|$ as a function of the Majorana phase $\beta$. These determine the ratios $b/a$ and $c/a$. Since we assume the lightest neutrino mass $m_1\approx0$, $\beta$ is the only Majorana phase. Lastly, also as a function of $\beta$, we will compute the two complex phases $\eta_2$ and $\eta_3$. We use rephasing freedom to choose $\eta_2$ and $\eta_3$ to enter into the Dirac neutrino mass matrix through elements $(m_{D})_{22}$ and $(m_{D})_{32}$ respectively --- with all other phases zero. We then supplement these calculations with $\chi^2$ analyses in order to guide the search for viable Dirac mass matrices. In all following calculations we compute the PMNS matrix elements to the second-order using the TBC3 ansatz in Eq.~(\ref{TBC3_ansatz}) along with Eqs.~(\ref{PMNS_1st_order})~and/or~(\ref{PMNS_2nd_order_PMNS_King_Corrections}).

\begin{figure}[htbp]
    \centering
    \includegraphics[width=1\linewidth]{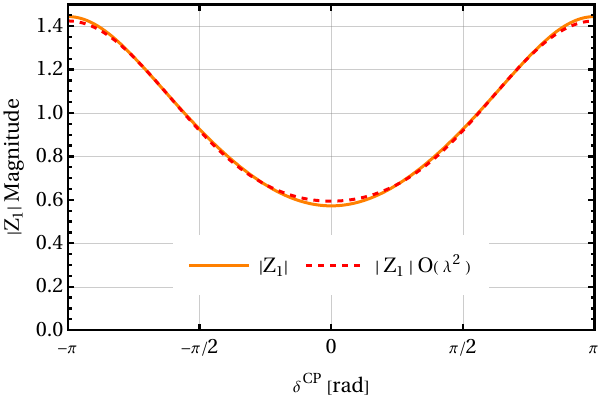}
    \caption{$|Z_1|=|\tilde{e}|/|\tilde{f}|$\: as a function of $\delta^{CP}$.}
    \label{fig:z1}
\end{figure}

In Figure~\ref{fig:z1} $|Z_1|$ and $|Z_1| \:\:\mathcal{O}(\lambda^2)$ are plotted as functions of $\delta^{CP}$. The $|Z_i|$ terms are obtained by substituting the second-order matrix elements directly into Eq.~(\ref{z3_explicit_expressions}) and then computing the results exactly. In contrast the $|Z_i|\:\:\mathcal{O}(\lambda^2)$ terms are truncated at second-order in $\lambda$ before the modulus is computed. Given that the PMNS matrix elements are second-order expansions in $\lambda$ themselves, the $|Z_i|\:\:\mathcal{O}(\lambda^2)$ terms are more theoretically sound. Plotting both approximations makes their structure clear and provides information on the stability of the approximations taken. We note that $\delta^{CP}=-\pi$ corresponds approximately to the simple identity $|Z_1|\:\:\mathcal{O}(\lambda^2)\approx|Z_1|\approx\sqrt{2}$.

\begin{figure}[htbp]
    \centering
    \includegraphics[width=1\linewidth]{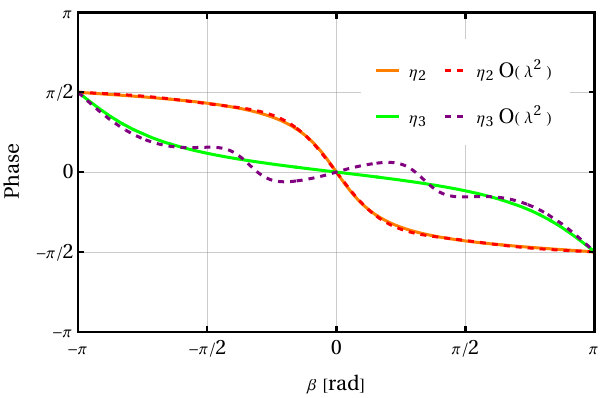}
    \caption{$\delta^{CP}=-\pi\,\mathrm{rad}$, $\eta_2$, $\eta_3$ as functions of the Majorana phase $\beta$.}
    \label{fig:etas - pi}
\end{figure}

\begin{figure}
    \centering
    \includegraphics[width=1\linewidth]{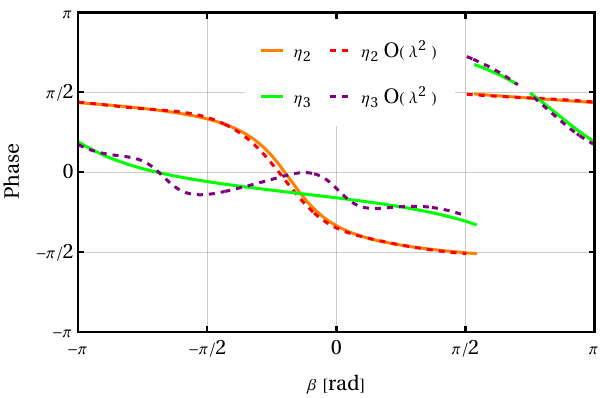}
    \caption{$\delta^{CP}=-5\pi/6\,\mathrm{rad}$, $\eta_2$, $\eta_3$ as functions of the Majorana phase $\beta$.}
    \label{fig:etas - 5pi6}
\end{figure}

Next, we compute $|Z_2|$ and $|Z_2| \:\:\mathcal{O}(\lambda^2)$ for $\delta^{CP}=-\pi,-5\pi/6$ in Figures~\ref{fig:z2,z3,-pi} and~\ref{fig:z2,z3,-5pi/6} as functions of $\beta$. The latter value of the CP-violating phase is chosen to be close to the NO best-fit in Eq.~(\ref{Eq:NuFIT_NO}) while employing only simple values. Then, for the same values of $\delta^{CP}$, we compute $\eta_2$ and $\eta_3$ also as functions of $\beta$. For the phases we employ the same two approximations employed with the $|Z_i|$. These are plotted in Figures~\ref{fig:etas - pi} and~\ref{fig:etas - 5pi6}. 

These plots can be used as a guide to find precise, simple, or otherwise viable Dirac textures that produce light neutrino observables in agreement with experimental data. In the remainder of this section we will list some such viable values --- both those that match data to high precision and those with simple values. We note that precise values, taken by reading points directly off of the curves in the plots, should in principle yield light-neutrino observable predictions closest to the TBC3 ansatz --- while being subject to error introduced by the approximations. Points read directly off the curves in this way will also correspond to a single Majorana phase. If we instead, in order to choose simple values, pick points that do not fall exactly on these curves, it is expected that the predictions of these values will deviate somewhat from the TBC3 ansatz and may in principle correspond to more than one Majorana phase. Since the primary interest in choosing VEVs is producing agreement with oscillation experiments in which the Majorana phases are unobservable we do not consider this further. 

In order to further guide the search we also perform a series of $\chi^2$ analyses. To discuss these, it must first be noted how precisely the Dirac neutrino mass matrix elements determine the light neutrino mass matrix. Defining the column vectors $A$ and $B$ via 
\begin{equation}
m_{\rm D}=\left( \begin{array}{cc}
0 & a    \\
e & b \cdot e^{i\eta_2}  \\
f & c \cdot e^{i\eta_3}
\end{array}
\right)=
\left( \begin{array}{cc}
A & B \\
\end{array}
\right),\label{dirac_mass_matrix_AB}
\end{equation}
and noting that in $m_{\mathrm{D},22}$ and $m_{\mathrm{D},32}$ that $e$ is the natural base while in $m_{\mathrm{D},21}$ it is the label for that entry, we then may write, after implementing the seesaw mechanism, the light-neutrino mass matrix as
\begin{equation}
m_\nu=m_a\bigg(AA^T+\epsilon_\nu BB^T\bigg).
\label{RSD_model_paramaters_light_neutrino_mass_matrix_as_function_of_A,B}
\end{equation}
Here $m_a$ controls the scale of the left-handed neutrino masses and has no effect on the mixing angles or CP-violating phase, and the desire for it to be purely real justifies the earlier choice of the sign convention of the seesaw mechanism in Eq.~(\ref{see_saw_mechanism_positive_right_habd_side}). Later it will be shown explicitly that $m_a$ depends on the RH neutrino mass $M=M_1=M_2$. In RSD, since the two RH neutrino masses are degenerate, it is the term $\epsilon_\nu$ which controls the ratio of the Yukawa couplings $y_{\mathrm{D}_2}/y_{\mathrm{D}_1}$ and is in turn responsible for the SD of the terms in Eq.~(\ref{SD1}). In conventional SD this role is shared between $\epsilon_\nu$ and the hierarchy of the RH neutrino masses.  

\begin{figure}[htbp]
    \centering
    \includegraphics[width=1\linewidth]{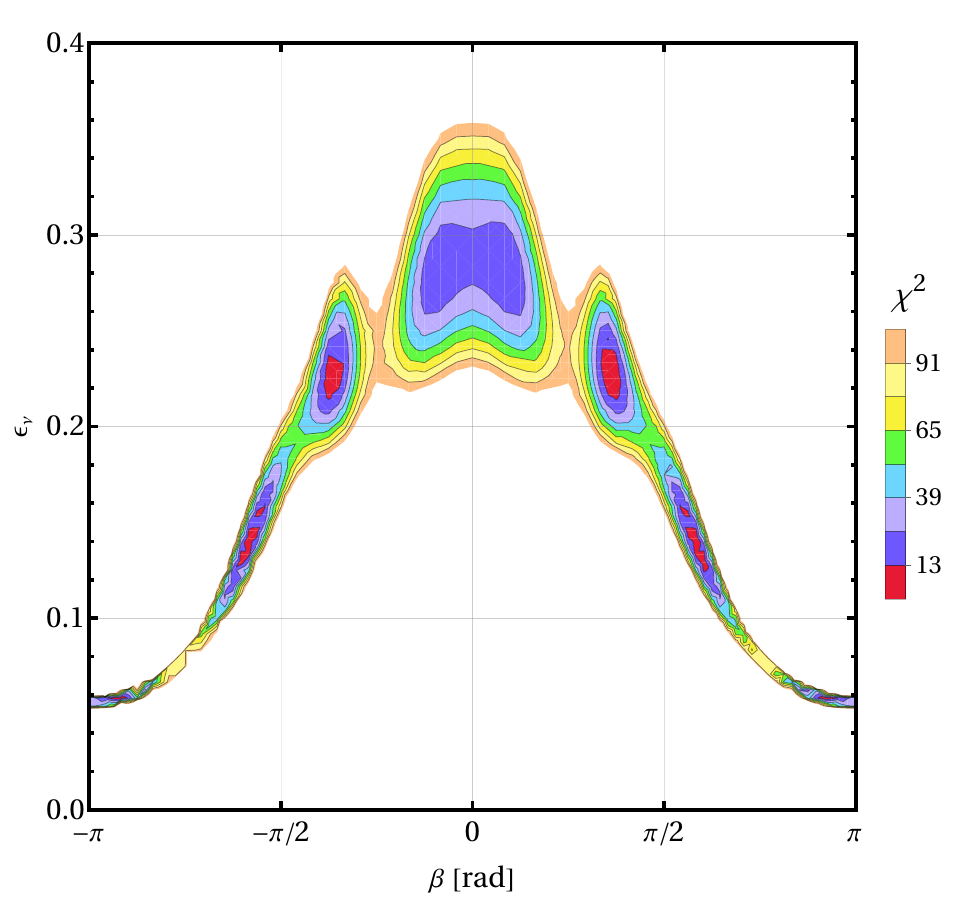}
    \caption{$\delta^{CP}=-\pi\,\mathrm{rad}$, reduced 3 degrees of freedom $\chi^2$ of $\theta_{ij}$, $\delta^{CP}$, $m_2/m_3$ as a function of both $\epsilon_\nu$ and the Majorana phase $\beta$. The unshaded region represents $\chi^2 > 105$.}
    \label{2d_reduced_chi_pi}
\end{figure}

\begin{figure}[htbp]
    \centering
    \includegraphics[width=1\linewidth]{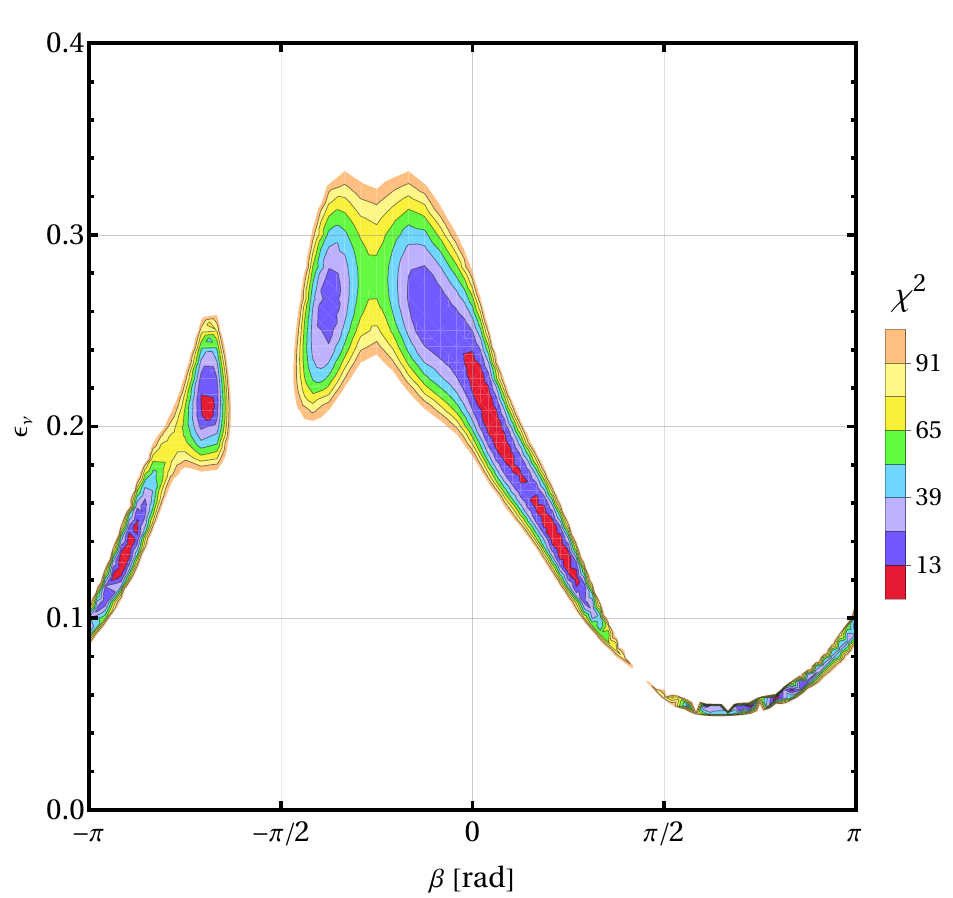}
    \caption{$\delta^{CP}=-5\pi/6\,\mathrm{rad}$, reduced 3 degrees of freedom $\chi^2$ of $\theta_{ij}$, $\delta^{CP}$, $m_2/m_3$ as a function of both $\epsilon_\nu$ and the Majorana phase $\beta$. The unshaded region represents $\chi^2 > 105$.}
    \label{2d_reduced_chi_5pi6}
\end{figure}

\begin{figure}[htbp]
    \centering
    \includegraphics[width=1\linewidth]{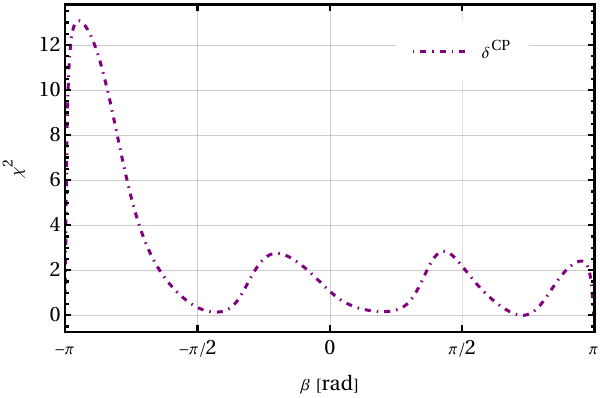}
    \caption{$\delta^{CP}=-\pi\,\mathrm{rad}$, single-degree-of-freedom $\chi^2$ of $\delta^{CP}$ for the minimizing value $\epsilon_\nu=0.1393$.}
    \label{fig:dcp_pi}
\end{figure}

\begin{figure}[htbp]
    \centering
    \includegraphics[width=1\linewidth]{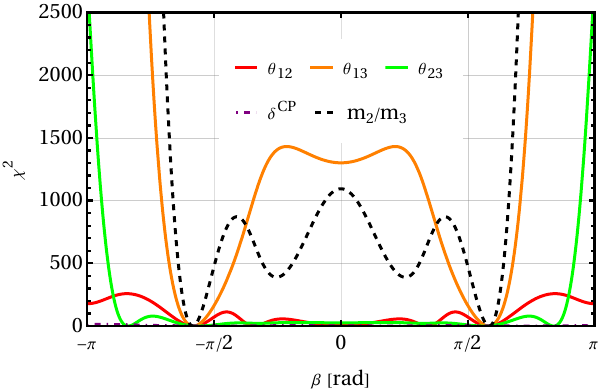}
    \caption{$\delta^{CP}=-\pi\,\mathrm{rad}$, individual $\chi^2$ for each of the 5 low-energy observables for the minimizing value $\epsilon_\nu=0.1393$.}
    \label{fig:all_observeables_pi}
\end{figure}

\begin{figure}
    \centering
    \includegraphics[width=1\linewidth]{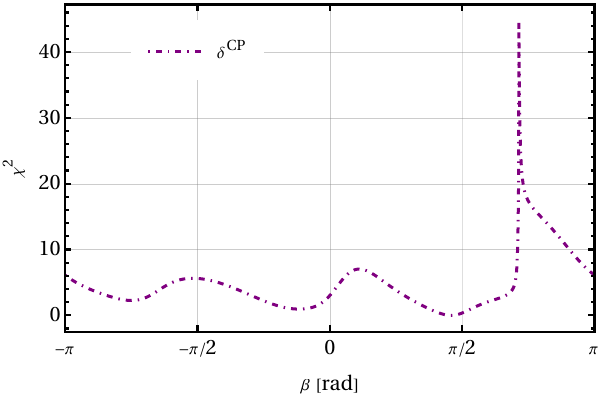}
     \caption{$\delta^{CP}=-5\pi/6\,\mathrm{rad}$, single degree of freedom $\chi^2$ of $\delta^{CP}$ for the minimizing value $\epsilon_\nu=0.1317$.}
    \label{fig:dcp_5pi6}
\end{figure}

\begin{figure}
    \centering
    \includegraphics[width=1\linewidth]{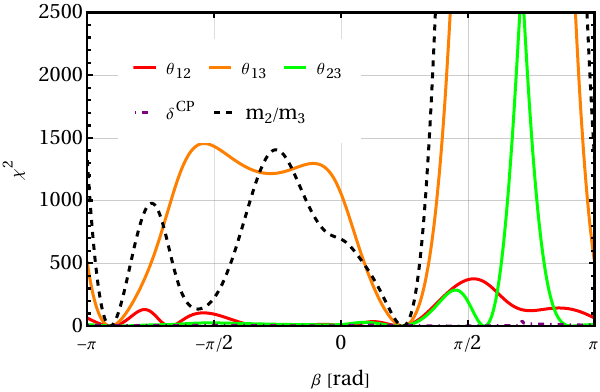}
    \caption{$\delta^{CP}=-5\pi/6\,\mathrm{rad}$, individual $\chi^2$ for each of the 5 low-energy observables for the minimizing value $\epsilon_\nu=0.1317$.}
    \label{fig:all5values_5pi6}
\end{figure}

In order to guide the survey of viable Dirac neutrino mass matrix elements we, for a particular choice of $\delta^{CP}$, use the Mixing Parameters Tools (MPT)~\cite{Mixing_Paramaters_Tools_Antusch:2005gp} Mathematica package to compute the mixing parameters from the light neutrino mass matrix derived from Eq.~(\ref{light_neutrino_mass_matrix_as_function_of_A,B}) as a function of the Majorana phase $\beta$. Using these predictions the reduced $\chi^2$ is plotted with three degrees of freedom of $\theta_{ij}$, $\delta^{CP}$, $m_2/m_3$ as a function of both $\beta$ and $\epsilon_\nu$. This is shown for $\delta^{CP}=-\pi,-5\pi/6$ in Figures~\ref{2d_reduced_chi_pi} and~\ref{2d_reduced_chi_5pi6}. In these two cases we then minimize the reduced $\chi^2$ to find the best-fit value of $\epsilon_\nu$. This best-fit value of $\epsilon_\nu$ is then used to compute the individual $\chi^2$ functions for each of the five mixing parameters independently as a function of $\beta$. This is shown for $\delta^{CP}=-\pi$ at the minimum value of $\epsilon_\nu\approx0.1393$ in Figures~\ref{fig:dcp_pi} and~\ref{fig:all_observeables_pi} then for $\delta^{CP}=-5\pi/6$,$\epsilon_\nu\approx0.1317$ in Figures~\ref{fig:dcp_5pi6} and~\ref{fig:all5values_5pi6}.

\begin{figure}[htbp]
    \centering
    \includegraphics[width=1\linewidth]{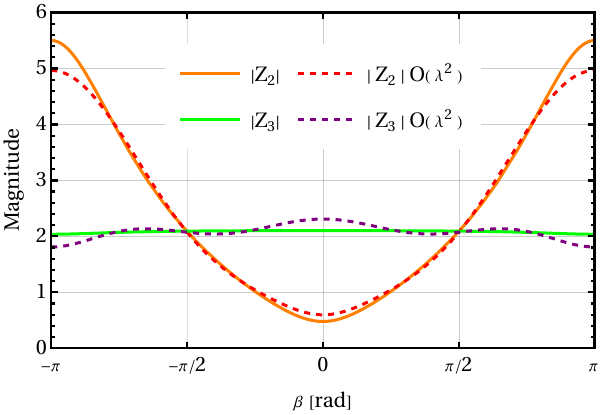}
    \caption{$\delta^{CP}=-\pi\,\mathrm{rad}$, $|Z_2|=|\tilde{b}|/|\tilde{a}|$, $|Z_3|=|\tilde{c}|/|\tilde{a}|$ as functions of the Majorana phase $\beta$.}
    \label{fig:z2,z3,-pi}
\end{figure}

\begin{figure}[htbp]
    \centering
    \includegraphics[width=1\linewidth]{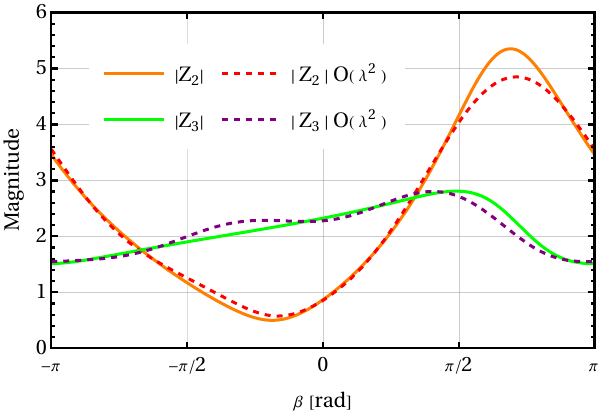}
    \caption{$\delta^{CP}=-5\pi/6\,\mathrm{rad}$, $|Z_2|=|\tilde{b}|/|\tilde{a}|$, $|Z_3|=|\tilde{c}|/|\tilde{a}|$ as functions of the Majorana phase $\beta$.}
     \label{fig:z2,z3,-5pi/6}
\end{figure}

Then, using the plots of the ratios $|Z_i|$ and complex phases $\eta_i$, along with the $\chi^2$ fits, we can search for viable values that produce good agreement with experimental data. Some such values are reported in Table~\ref{table_1_vevs}. The bold rows represent the best-fit parameters found purely via minimizing the reduced $\chi^2(\delta^{CP},\epsilon_\nu,\beta)$. For example, the alignment corresponding to $\chi^2=0.997$ is
\begin{equation}
\left( \begin{array}{cc}
A & B \\
\end{array}
\right)=\left( \begin{array}{cc}
0 & 1    \\
1.424 & 2.614\cdot e^{-i1.403}  \\
1 & 2.118\cdot e^{-i0.483}
\end{array}
\right),\:\:\:\:\epsilon_\nu=0.1393
\end{equation}
which via MPT yields 
\begin{equation}
    \theta_{12}=33.61^\circ,\:\:\:\:\theta_{13}=8.64^\circ,\:\:\:\:\theta_{23}=42.18^\circ
\end{equation}
which matches well with the TBC3 ansatz in Eq.~($\ref{TBC3_ansatz}$). We note here that $\theta_{23}$ falls outside the NuFIT $1\sigma$ allowed region --- a direct consequence of the equality assigned to the atmospheric and solar deviations in the TBC3 ansatz. 

Since only the bold best-fit values correspond to exact points on the curves $|Zi|(\beta),\:\eta_i(\beta)$, we do not list $\beta$ in Table~\ref{table_1_vevs}. The remaining values in the table have been found by examining both the best-fit points and the plots of $|Z_i|(\beta),\:\eta_i(\beta)$, and do not strictly fall on the curves. Looking for viable points in this way we found excellent agreement with experimental data, corresponding to $\chi^2=0.632$, for the values 
\begin{equation}
\left( \begin{array}{cc}
A & B \\
\end{array}
\right)=\left( \begin{array}{cc}
0 & 1    \\
\sqrt{2} & 1\cdot e^{-i\pi/3}  \\
1 & 2
\end{array}
\right),\:\:\:\:\epsilon_\nu=0.2758
\label{Best_VEVs}
\end{equation}
which has the appealing feature of $\eta_1=\eta_3=0$. This leads to 
\begin{equation}
    \theta_{12}=33.95^\circ,\:\:\:\:\theta_{13}=8.57^\circ,\:\:\:\:\theta_{23}=42.55^\circ
\end{equation}
along with $\delta^{CP}=192.2^\circ$ and $m_2/m_3=0.174$. Noting that per the NuFIT data in Eq.~\ref{Eq:NuFIT_NO}, in the NO case, the $1\sigma$ range of $m_2/m_3$ is $\sim 0.171-0.175$, the values predicted by this set of Dirac neutrino mass matrix elements are all in $1\sigma$ agreement with experimental data. In particular, good agreement is achieved by setting the remaining free parameter $m_a$ to
\begin{equation}
    m_a=0.01315\,\mathrm{eV}
    \label{m_a_value}
\end{equation}
corresponding to $\Delta m^2_{21}=7.54\times10^{-5}\,\mathrm{eV^2}$ and $\Delta m^2_{31}=2.48\times10^{-3}\,\mathrm{eV^2}$. We note that $\theta_{23}$ is only marginally within the NuFIT $1\sigma$ allowed region as stated in Eq.~(\ref{Eq:NuFIT_NO}). This is again a direct consequence of the TBC3 ansatz and the equality of the solar and atmospheric deviations assumed therein. Throughout the rest of this work this alignment is referred to, along with its corresponding set of $\eta_2$, $\eta_3$, $\epsilon_\nu$, $m_a$ values, as the Simple $(\sqrt{2},1;1,2)$ Alignment. 

\begin{table*}[htbp]
\caption{Inputs: Flavon alignments/VEVs, phases, and corresponding minima of $\epsilon_\nu$ derived from the Master Formula calculations and $\chi^2$ analysis under the TBC3 ansatz. Outputs: Low-energy observables computed via Mixing Parameters Tools~\cite{Mixing_Paramaters_Tools_Antusch:2005gp} using the Dirac mass matrix constructed from the alignments given in ``Inputs" and corresponding reduced $\chi^2$ value. Bold rows indicate reduced $\chi^2(\delta^{CP},\epsilon_\nu,\beta)$ best-fit points. The best alignment has its $\chi^2$ value, 0.632, marked with stars. }\label{table_1_vevs}
\normalsize
\setlength{\tabcolsep}{4pt}
\renewcommand{\arraystretch}{1.2}
\begin{tabular}{|c|c|c|c|c|c|c|c|c|c|c|c|}
\hline
\multicolumn{6}{|c|}{\textbf{Inputs }(\textbf{$\boldsymbol{\delta^{CP}},\boldsymbol{\eta_i}$)\,[rad]}} & \multicolumn{6}{c|}{\textbf{Outputs }(\textbf{$\boldsymbol{\theta_{ij},}\boldsymbol{\delta^{CP}})$[$^\circ$]}} \\[0.2cm]
\hline
\textbf{$\boldsymbol{\delta^{CP}}$} & \textbf{$A=(0,e,f)^T$} & \textbf{$B=(1,b,c)^T$} & \textbf{$\boldsymbol{\eta_2}$} & \textbf{$\boldsymbol{\eta_3}$} & \textbf{$\boldsymbol{\epsilon_\nu}$} & \textbf{$\boldsymbol{\delta^{CP}}$} & \textbf{$\boldsymbol{\theta_{12}}$} & \textbf{$\boldsymbol{\theta_{13}}$} & \textbf{$\boldsymbol{\theta_{23}}$} & \textbf{$m_2/m_3$} & \textbf{Reduced $\boldsymbol{\chi^2}$} \\[0.2cm]
\hline
$\mathbf{\text{-}\boldsymbol{\pi}}$ & $\mathbf{1.424, 1}$ & $\mathbf{2.614, 2.118}$ & $\mathbf{\text{-}1.403}$ & $\mathbf{\text{-}0.483}$ & $\mathbf{0.1393}$ & $\mathbf{180.5}$ & $\mathbf{33.61}$ & $\mathbf{8.64}$ & $\mathbf{42.18}$ & $\mathbf{0.172}$ & $\mathbf{0.997}$ \\[0.2cm]
\hline
$\text{-}\pi$ & $\sqrt{2}, 1$ & $\frac{3}{2}\sqrt{3}, \frac{3}{2}\sqrt{2}$ & $\text{-}11\pi/24$ & $\text{-}\pi/6$ & $0.1451$ & $174.6$ & $33.78$ & $8.42$ & $42.44$ & $0.175$ & $2.513$ \\[0.2cm]
$\text{-}\pi$ & $\sqrt{2}, 1$ & $1, 2$ & $\text{-}\pi/3$ & $0$ & $0.2758$ & $192.2$ & $33.95$ & $8.57$ & $42.55$ & $0.174$ & $^\star\:0.632\:^\star$ \\[0.2cm]
\hline
$\mathbf{\text{-}5\boldsymbol{\pi}/6}$ & $\mathbf{1.346, 1}$ & $\mathbf{2.928, 1.568}$ & $\mathbf{1.33}$ & $\mathbf{0.376}$ & $\mathbf{0.1317}$ & $\mathbf{150.0}$ & $\mathbf{33.69}$ & $\mathbf{8.63}$ & $\mathbf{42.44}$ & $\mathbf{0.173}$ & $\mathbf{1.677}$ \\[0.2cm]
\hline
$\text{-}5\pi/6$ & $\sqrt{2}, 1$ & $2\sqrt{2}, \sqrt{3}$ & $7\pi/16$ & $\pi/8$ & $0.1369$ & $163.3$ & $33.68$ & $8.47$ & $42.80$ & $0.175$ & $1.910$ \\[0.3cm]
$\text{-}5\pi/6$ &$4/3,1$& $3,3/2$ &$5\pi/12$ &$\pi/8$ &$0.1303$ &$144.61$ &$33.95$ &$8.77$ &$42.77$ &$0.171$ &$3.131$ \\[0.2cm]
\hline
\end{tabular}
\end{table*}

\section{Lagrangian}
\label{sec:Lagrangian}
In this section the field contents and charge assignments of an explicit $A_4\times (Z_3)^4\times(Z_2)^2$-symmetric Lagrangian are presented, which, together with the flavon alignments found in the previous section, may be used to realize both the experimental values of the low-energy observables while simultaneously admitting resonant leptogenesis. The approach taken here is similar to those taken in works such as Refs.~\cite{Esteban2024JHEP,PhysRevD.85.031903,TM2_nonresonant_He_2007,King_nice_Vev_search_Master_formula_paper_king2013minimalpredictiveseesawmodel,King2011TrimaximalNM,Thapa_2021} and most notably Refs.~\cite{Altarelli2005A4,Ma2001A4}. Before the observation of a non-zero reactor angle and measurement of $\sin^2\theta_{12}<1/3$, the previously favored neutrino mixing ansatz was TBM mixing. Extensions of the SM based on an $S_4$ family symmetry often naturally reproduce TBM mixing and so this is a widely studied framework~\cite{King_Review}.  

$S_4$, the symmetry group of all permutations of four objects, may be identified with the symmetry of a cube. $S_4$ contains the subgroup $A_4$ comprising all even permutations of four objects --- this subgroup may be identified with the symmetry of a tetrahedron. We will construct a model whose unbroken phase is $A_4$-symmetric. This is supplemented with several shaping symmetries forbidding certain undesired Yukawa and Majorana sector operators.

$A_4$ is chosen as the underlying family symmetry because simple $A_4$ constructions naturally yield TM$_2$ mixing~\cite{CS_Lam_Group_theory_and_dynamics_of_neutrino_mixing}, which accommodates a non-zero reactor angle but is no longer compatible with the observed solar mixing angle. Since the main focus is the neutral-lepton sector, the FN mechanism~\cite{FROGGATT1979277} is assumed to generate the observed charged-lepton mass hierarchy in a typical manner and we do not specify the FN fields or charge assignments explicitly. For numerical calculations we take the charged-lepton mass matrix eigenvalues to be equal to their best-fit values as per the PDG\cite{PDG_PhysRevD.110.030001}.

The singlets of $A_4$ are $\mathbf{1}$, $\mathbf{1'}$, and $\mathbf{1''}$. These may be identified with 1, $\omega$, $\omega^2$, respectively, where $\omega$ is the cubic root of unity\cite{Ishimori_2010}. This leads to the following tensor product rules for the $A_4$ singlets
\begin{equation}    
\mathbf{1}\otimes\mathbf{1}=\mathbf{1},\:\:\:\:\:\:\:\:\mathbf{1'}\otimes\mathbf{1'}=\mathbf{1''},\:\:\:\:\:\:\:\:\mathbf{1''}\otimes\mathbf{1''}=\mathbf{1'},\:\:\:\:\:\:\:\:\mathbf{1'}\otimes\mathbf{1''}=\mathbf{1}.
\end{equation}
For triplets, the product rules are
\begin{equation}    
\mathbf{1'}\otimes\mathbf{3}=\mathbf{3},\:\:\:\:\:\:\:\:\mathbf{1''}\otimes\mathbf{3}=\mathbf{3},\:\:\:\:\:\:\:\:\mathbf{3}\otimes\mathbf{3}=\mathbf{1}\oplus\mathbf{1'}\oplus\mathbf{1''}\oplus\mathbf{3}\oplus\mathbf{3}.
\end{equation}
For arbitrary vectors $\mathbf{A}$ and $\mathbf{B}$ in an $\mathbb{R}^3$ dimensional representation of $A_4$, the triplet product $\mathbf{3}\otimes\mathbf{3}$ may be written explicitly as \cite{Ishimori_2010}
\begin{align}
(\mathbf{A})_{\mathbf{3}} \times (\mathbf{B})_{\mathbf{3}} &= (A_x B_x + A_y B_y + A_z B_z)_{\mathbf{1}} \notag\\
&\quad + (A_x B_x + \omega A_y B_y + \omega^2 A_z B_z)_{\mathbf{1}'} \notag\\
&\quad + (A_x B_x + \omega^2 A_y B_y + \omega A_z B_z)_{\mathbf{1}''} \notag\\
&\quad + \begin{pmatrix} A_y B_z + A_z B_y \\ A_z B_x + A_x B_z \\ A_x B_y + A_y B_x \end{pmatrix}_{\mathbf{3}}\notag \\&\quad+ \begin{pmatrix} A_y B_z - A_z B_y \\ A_z B_x - A_x B_z \\ A_x B_y - A_y B_x \end{pmatrix}_{\mathbf{3}}.
\label{a4_triplet_tensor_products}
\end{align}
\begin{table*}[htbp]
    \centering
    \renewcommand{\arraystretch}{1.6}
    \caption{$A_4\times (Z_3)^4\times(Z_2)^2$-symmetric Lagrangian field contents and representations under $A_4$ and the auxiliary $Z_3^X$ and $Z_2^X$ shaping symmetries. Here $\omega$ is the cubic root of unity, $\omega=e^{2i\pi/3}$.}
    \label{field content}
    \vspace{1em}
    \begin{tabular*}{\textwidth}{@{\extracolsep{\fill}}|l@{\hspace{2.5em}}|*{15}{c}|}
        \hline
        \textit{}  &\textbf{$L_l$} &H& \textbf{${\varphi}$} & \textbf{$e_R$} & \textbf{$\mu_R$} & \textbf{$\tau_R$} & \textbf{$N_1$} & \textbf{$N_2$} & \textbf{$N_3$} & \textbf{$\phi_1$} & \textbf{$\phi_2$} & \textbf{$\phi_3$} & \textbf{$\chi_1$} & \textbf{$\chi_2$} & \textbf{$\chi_3$} \\ 
        \hline
        $A_4$& $\mathbf{3}$ &$\mathbf{1}$& $\mathbf{3}$ & $\mathbf{1}$& $\mathbf{1'}$ & $\mathbf{1''}$ & $\mathbf{1}$ & $\mathbf{1}$ & $\mathbf{1'}$ & $\mathbf{3}$ & $\mathbf{3}$ & $\mathbf{3}$ & $\mathbf{1}$ & $\mathbf{1}$ & $\mathbf{1''}$ \\ 
        \hline 
        $Z_3^{\scriptscriptstyle N_1}$ & 1 & 1 & 1 & 1 & 1 & 1 & $\omega$ & 1 & 1 & $\omega^2$ & 1 & 1 & $\omega^2$ & 1 & 1 \\ 
        $Z_3^{\scriptscriptstyle N_2}$ & 1 & 1 & 1 & 1 & 1 & 1 & 1 & $\omega$ & 1 & 1 & $\omega^2$ & 1 & 1 & $\omega^2$ & 1 \\ 
        $Z_3^{\scriptscriptstyle N_3}$ & 1 & 1 & 1 & 1 & 1 & 1 & 1 & 1 & $\omega$ & 1 & 1 & $\omega^2$ & 1 & 1 & $\omega^2$ \\ 
        \hline
         $Z_3^{\scriptscriptstyle\phi/\chi}$ &$\omega^2$ & 1 & 1 & $\omega^2$ & $\omega^2$ & $\omega^2$ & $\omega$ & $\omega$ & $\omega$ & $\omega$ & $\omega$ & $\omega$ & $\omega^2$ & $\omega^2$ & $\omega^2$ \\ 
        \hline 
        $Z_2^{\scriptscriptstyle N_1}$ & 1 & 1 & 1 & 1 & 1 & 1 & $-1$ & 1 & 1 & $-1$ & 1 & 1 & 1 & 1 & 1 \\ 
        $Z_2^{\scriptscriptstyle N_2}$ & 1 & 1 & 1 & 1 & 1 & 1 & 1 & $-1$ & 1 & 1 & $-1$ & 1 & 1 & 1 & 1 \\ 
        \hline 
    \end{tabular*}
\end{table*}
Assigning the fields to the symmetries in Table~\ref{field content} yields
%
\begin{align}\label{Lagrangian}
    -\mathscr{L}\supset&\dfrac{y_e}{\Lambda}(\bar{L}_lH\varphi)_1e_R+\dfrac{y_\mu}{\Lambda}(\bar{L}_lH\varphi)_{1''}\mu_R+\dfrac{y_\tau}{\Lambda}(\bar{L}_lH\varphi)_{1'}\tau_R\notag\\
    +&\dfrac{y_{\mathrm{D}_1}}{\Lambda}(\bar{L}_l\tilde{H}\phi_1)_{1}N_1+\dfrac{y_{\mathrm{D}_2}}{\Lambda}(\bar{L}_l\tilde{H}\phi_2)_{1}N_2\\\notag
    +&\dfrac{y_{\mathrm{D}_3}}{\Lambda}(\bar{L}_l\tilde{H}\phi_3)_{1''}N_3\\\notag
    +&\dfrac{1}{2}\sum_{i=1}^3(\bar{N}_i^cN_i)\biggr[z_{i,1}\chi^*_i+z_{i,2}\dfrac{\chi_i \chi_i}{\Lambda}\biggr] +\mathrm{h.c.}, 
\end{align}
%
where $L_l$ is the triplet of left-handed lepton doublets, $H$ is the Higgs doublet, and $\tilde{H}=i\sigma_2H^*$ with $\sigma_2$ being the second Pauli matrix.

The $Z_3^X$ and $Z_2^X$ shaping symmetries and associated charges are chosen such that these are the only allowed four- or five-dimensional lepton sector Yukawa or Majorana operators which may be constructed from the fields in Table~\ref{field content}. That is, they prevent unwanted off-diagonal mass terms such as $(\bar{N}_1N_2^c)_{1}\chi_1$, and $(\bar{N}_1N_2^c)_{1}\phi_1\phi_2$ and enforce a direct correspondence between the RH neutrinos, the flavons, and the singlets. Furthermore, they ensure the strict diagonality of the Majorana mass matrix. This is crucial in order to maintain the SD relationship in Eq.~(\ref{SD1}).

In the unbroken phase this model respects the full $A_4=\{F,G_2\}$ symmetry and all of the shaping symmetries. Here $G_2$ is the generator of $S_4=\{F,G_1,G_2,G_3\}$ which may be used to produce TM$_2$ mixing~\cite{CS_Lam_Group_theory_and_dynamics_of_neutrino_mixing,King_Review}. After the flavons $\varphi,\phi_i$ develop VEVs below the flavor symmetry breaking scale the symmetry is broken such that the charged-lepton sector respects $F$ independently, and the neutral-lepton sector has its residual symmetry completely broken. In the basis where the charged-lepton mass matrix is diagonal, the necessary VEV of $\varphi$ is $\braket{\varphi}=(1,0,0)^{T}$.

Also imposed on this Lagrangian is the requirement that the fields $N_1$ and $N_2$ belong to the trivial singlet representation of $A_4$, as opposed to $\mathbf{1'}$ or $\mathbf{1''}$. This ensures that the VEVs enter into the Lagrangian exactly as defined in Eq.~(\ref{dirac_mass_matrix_AB}). Were the non-trivial singlets to be used, additional rephasing would be required in order for the VEVs to enter into the Dirac mass matrix as desired. 

Since $N_3$ is taken to be much heavier than the leptogenesis scale, it decouples from any of the regions of interest and no details about the structure of $\braket{\phi_3}$ are assumed. Since we require for $M_1=M_2$ we assume that the VEVs of the scalar fields $\chi_i$ and couplings $z_{i,j}$ satisfy
\begin{equation}
\braket{\chi_1}(z_{1,1}+z_{1,2})=\braket{\chi_2}(z_{2,1}+z_{2,2})\ll\braket{\chi_3}(z_{3,1}+z_{3,2})    
\end{equation}
 yielding $M_1=M_2\ll M_3$.  

 With the Lagrangian established in Eq.~(\ref{Lagrangian}) which may be used to realize RSD with the VEV alignments presented in the previous section, we now proceed to investigate whether such a Lagrangian admits resonant leptogenesis. 
    
\section{Leptogenesis}
\label{sec:Leptogenesis}
\subsection{Connecting RSD Parameters to Lagrangian Terms}
Before computing the baryon asymmetry predicted by our model, we first relate the RSD model parameters of Eq.~(\ref{RSD_model_paramaters_light_neutrino_mass_matrix_as_function_of_A,B}) directly to the parameters of the Lagrangian in Eq.~(\ref{Lagrangian}). After EW symmetry breaking, and after integrating out the ultra-heavy RH neutrino $N_3$, the seesaw mechanism yields 
\begin{align}
m_\nu&=m_{\rm D}M_{\rm R}^{-1}m_{\rm D}^T\notag\\&=\dfrac{v^2}{2M\Lambda^2}\biggr(y^2_{\mathrm{D}_1}\braket{\phi_1}\braket{\phi_1}^T+y^2_{\mathrm{D}_2}\braket{\phi_2}\braket{\phi_2}^T\biggr),\label{m_nu_in_Lagrangian_parameters}
\end{align}
where $v=246$\,GeV is the Higgs VEV. In terms of the column vectors $A$, $B$, we restate Eq.~(\ref{RSD_model_paramaters_light_neutrino_mass_matrix_as_function_of_A,B}) 
\begin{equation}
m_\nu=m_a\bigg(AA^T+\epsilon_\nu BB^T\bigg).
\label{light_neutrino_mass_matrix_as_function_of_A,B}
\end{equation}
Comparing Eq.~(\ref{m_nu_in_Lagrangian_parameters}) to Eq.~(\ref{light_neutrino_mass_matrix_as_function_of_A,B}) immediately yields
\begin{equation}
    m_a=\dfrac{v^2y^2_{\mathrm{D}_1}}{2M\Lambda^2},\:\:\:\: \epsilon_\nu=\dfrac{y^2_{\mathrm{D}_2}}{y^2_{\mathrm{D}_1}}.
    \label{m_a_LAMBDA_y_d1_constaint_equation}
\end{equation}
Crucially, this relates the heavy RH neutrino mass scale, $M$, to the cutoff scale $\Lambda$ and the Yukawa couplings. Once $m_a$ is fixed by the low-energy observables and any one of the terms $\Lambda$, $y_{\mathrm{D}_1}$, $M$ is chosen, the remaining two quantities on the RHS of Eq.~(\ref{m_a_LAMBDA_y_d1_constaint_equation}) are no longer independent quantities. Rearranging Eq.~(\ref{m_a_LAMBDA_y_d1_constaint_equation}) yields the useful relationship
\begin{equation}
    M=\dfrac{v^2}{2m_a}\biggr(\dfrac{y_{\mathrm{D}_1}}{\Lambda}\biggr)^2.
    \label{majorana_mass_in_terms_of_LAMBDA_etc}
\end{equation}

Given that $\Lambda$ must be sufficiently large for the flavons to develop VEVs in the leptogenesis regime, and $y_{\mathrm{D}_1}$ is typically required to be $\mathcal{O}(1)$ or less, this sets strict boundaries on the available parameter space. Later in this section we will determine how leptogenesis proceeds within these bounds. Furthermore, assuming the positive root of $\epsilon_\nu$, the Yukawa matrix can now be written as 

\begin{equation}
    Y=\dfrac{1}{\Lambda}\begin{pmatrix}
    y_{\mathrm{D}_1}A&y_{\mathrm{D}_2}B
    \end{pmatrix}=\dfrac{y_{\mathrm{D}_1}}{\Lambda}\begin{pmatrix}
        \braket{\phi_{1,1}}& \epsilon_\nu^{\frac{1}{2}}\braket{\phi_{2,1}}\\
        \braket{\phi_{1,2}}& \epsilon_\nu^{\frac{1}{2}}\braket{\phi_{2,2}}\\
        \braket{\phi_{1,3}}& \epsilon_\nu^{\frac{1}{2}}\braket{\phi_{2,3}}\\
    \end{pmatrix}.
\end{equation}

Thus, a one-to-one correspondence between the phenomenological RSD parameters and the Lagrangian parameters is established, and the relationships between them, once a particular $m_a$ is chosen, are demonstrated.

\subsection{Breaking RH Neutrino Mass Degeneracy}
The primary condition for resonant leptogenesis is for the exact degeneracy of the two heavy RH neutrinos to be slightly broken by an amount proportional to half the RH neutrinos' decay widths. Recalling that several symmetries have been imposed in order to keep the Majorana mass matrix strictly diagonal for terms up to dimension five in the Lagrangian, it is assumed that the small off-diagonal corrections arise from higher-dimensional operators present at the scale of leptogenesis. We take these to introduce a splitting between the Majorana masses proportional to $\Delta\ll M$ like 

\begin{equation}
M_{\rm R}=
\left( \begin{array}{cc}
M & 0    \\
0 & M  \\
\end{array}\right)
\xrightarrow{\mathrm{Above\:\:cutoff}}
M^l_{\rm R}=\left( \begin{array}{ccc}
M & \Delta     \\
\Delta & M \\
\end{array}\right).
\end{equation}

In leptogenesis computations one must track the evolution of the mass eigenstates and so the mass matrix is rendered diagonal so that

\begin{equation}
\Omega M_{\rm R}^l \Omega^T=\left( \begin{array}{ccc}
M-\Delta  & 0    \\
0 & M+\Delta  \\
\end{array}\right)
\label{diagonalized_majorana_mass_Matrix_with_mass_splitting}
\end{equation}
where $\Omega$ is 
\begin{equation}
    \Omega=\dfrac{1}{\sqrt{2}}\begin{pmatrix}
    -1&1\\
    1&1
    \end{pmatrix}.
\end{equation}
This leads to the following Yukawa matrix in the basis where the RH neutrino mass matrix is now diagonal
\begin{align}
    Y\Omega&=Y^l\notag\\
    &=\dfrac{y_{\mathrm{D}_1}}{\sqrt{2}\Lambda}\begin{pmatrix}
     \epsilon_\nu^{\frac{1}{2}}\braket{\phi_{2,1}}& \epsilon_\nu^{\frac{1}{2}}\braket{\phi_{2,1}}\\
     -\braket{\phi_{1,2}}+\epsilon_\nu^{\frac{1}{2}}\braket{\phi_{2,2}}&\braket{\phi_{1,2}}+\epsilon_\nu^{\frac{1}{2}}\braket{\phi_{2,2}}\\
     -\braket{\phi_{1,3}}+\epsilon_\nu^{\frac{1}{2}}\braket{\phi_{2,3}}&\braket{\phi_{1,3}}+\epsilon_\nu^{\frac{1}{2}}\braket{\phi_{2,3}}\\
    \end{pmatrix}
\end{align}
where the same texture-zero condition used elsewhere in this work is imposed: $\braket{\phi_{1,1}}=0$.

\subsection{Resonant Leptogenesis}

\begin{figure*}[htbp]
\centering
\begin{tikzpicture}[scale=.72]

\begin{scope}[shift={(0,0)}]

\coordinate (L) at (0,0);
\coordinate (X) at (1.62,0);
\coordinate (V) at (3.05,0);

\draw[ferm] (L)--(X);
\draw[ferm] (V)--(X);

\draw[mass] ($(X)+(-0.11,-0.11)$)--($(X)+(0.11,0.11)$);
\draw[mass] ($(X)+(-0.11,0.11)$)--($(X)+(0.11,-0.11)$);

\coordinate (OutH) at ($(V)+(1.05,0.72)$);
\coordinate (OutL) at ($(V)+(1.05,-0.72)$);

\draw[scalar] (OutH)--(V);
\draw[ferm]   (OutL)--(V);

\node at (.81,.35){$N_i$};
\node at (2.35,.35){$N_i$};

\node[right] at (OutH){$\mathbf{H}^\dagger$};
\node[right] at (OutL){$L^C$};

\node at (1.52,-1.65){(a)};

\end{scope}

\begin{scope}[shift={(6.1,0)}]

\coordinate (A) at (0,0);
\coordinate (B) at (1.5,0);
\coordinate (X) at (3.6,0);
\coordinate (V) at (4.8,0);

\draw[ferm] (A)--(B);
\draw[plain] (X)--(V);

\draw[scalar mid]
(B)..controls +(0,1.0) and +(0,1.0)..
node[above=2pt]{$\mathbf{H}$}(X);

\draw[ferm] (B)-- node[below]{$L$} (X);

\coordinate (M) at ($(X)!0.5!(V)$);
\draw[mass] ($(M)+(-0.11,-0.11)$)--($(M)+(0.11,0.11)$);
\draw[mass] ($(M)+(-0.11,0.11)$)--($(M)+(0.11,-0.11)$);

\coordinate (OutHb) at ($(V)+(1.05,0.72)$);
\coordinate (OutLb) at ($(V)+(1.05,-.72)$);

\draw[scalar] (OutHb)--(V);
\draw[ferm]   (OutLb)--(V);

\node at (.75,.45){$N_i$};
\node at (4.2,.45){$N_j$};

\node[right] at (OutHb){$\mathbf{H}^\dagger$};
\node[right] at (OutLb){$L^C$};

\node at (2.55,-1.65){(b)};

\end{scope}

\begin{scope}[shift={(14.1,0)}]

\coordinate (A) at (0,0);
\coordinate (V1) at (1.2,0);
\coordinate (TR) at (2.7,0);
\coordinate (BR) at (2.7,-1.2);
\coordinate (OutHc) at (3.75,0.78);
\coordinate (OutLc) at (3.95,-1.2);

\draw[ferm] (A)--(V1);
\node at (0.6,-0.35) {$N_i$};

\draw[ferm] (V1)--(TR);
\node at (1.95,0.38) {$L$};

\draw[scalar] (V1)--(BR);
\node at (1.55,-0.75) {$\mathbf{H}$};

\draw[scalar] (OutHc)--(TR);
\node[right] at (OutHc){$\mathbf{H}^\dagger$};

\draw[plain] (TR)--(BR);
\coordinate (M) at (2.7,-0.6);
\draw[mass] ($(M)+(-0.11,-0.11)$)--($(M)+(0.11,0.11)$);
\draw[mass] ($(M)+(-0.11,0.11)$)--($(M)+(0.11,-0.11)$);
\node at (3.15,-0.6) {$N_j$};

\draw[ferm] (OutLc)--(BR);
\node[right] at (OutLc){$L^C$};

\node at (1.95,-1.85){(c)};

\end{scope}

\end{tikzpicture}
\caption{Feynman diagrams describing the tree (a) and one-loop self-energy (b) and one-loop vertex decays (c) of RH Majorana neutrinos. A small mass splitting of the RH masses leads to the resonant enhancement of diagram (b) which generates the large CP asymmetry responsible for resonant leptogenesis as outlined in Ref.~\cite{Resonant_lepto_Pilaftsis_2004}.}
\label{fig:feynmamn}
\end{figure*}
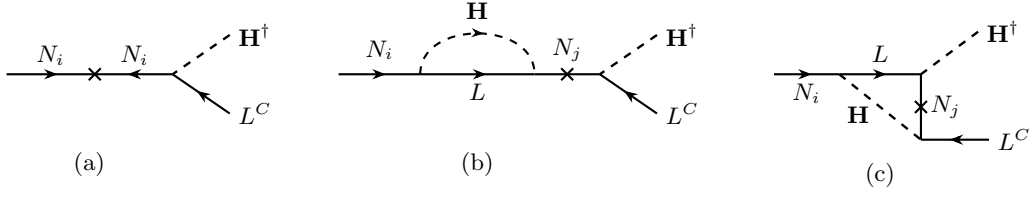

In order for resonant leptogenesis to occur, as pointed out by the seminal work of Pilaftsis and Underwood in Ref.~\cite{Resonant_lepto_Pilaftsis_2004}, the following conditions must be met
\begin{equation}
m_{N_i} - m_{N_j} \sim \frac{\Gamma_{N_{i,j}}}{2}, \quad \frac{\lvert \text{Im}(Y^{l\dagger}Y^l)^2_{ij} \rvert}{(Y^{l\dagger}Y^l)_{ii}(Y^{l\dagger}Y^l)_{jj}} \sim 1,\label{resonance_conditions}
\end{equation}
where $m_{N_i}$ are the RH neutrino masses in the basis where the Majorana mass matrix is diagonal, i.e., the eigenvalues of Eq.~(\ref{diagonalized_majorana_mass_Matrix_with_mass_splitting}). The tree-level decay widths relevant to Eq.~(\ref{resonance_conditions}) and pertaining to Figure~\ref{fig:feynmamn}(a) are 
\begin{equation}
\quad \Gamma^{(0)}_{N_i} = \frac{(Y^{l\dagger}Y^l)_{ii}}{8\pi} m_{N_i}.
\end{equation}
For the duration of this section the same notation as Ref.~\cite{Resonant_lepto_Pilaftsis_2004} is employed. The CP-violating quantities that drive the lepton asymmetry for the two RH neutrinos, $\delta_{N_{1,2}}$, are given by 
\begin{align}
\delta_{N_i} &= \frac{\Gamma(N_i \rightarrow LH) - \Gamma(N_i \rightarrow L^CH^\dagger)}{\Gamma(N_i \rightarrow LH) + \Gamma(N_i \rightarrow L^CH^\dagger)} \notag\\&= \frac{(\bar{Y}^{l\dagger}_+ \bar{Y}^l_+)_{ii} - (\bar{Y}^{l\dagger}_- \bar{Y}^l_-)_{ii}}{(\bar{Y}^{l\dagger}_+ \bar{Y}^l_+)_{ii} + (\bar{Y}^{l\dagger}_- \bar{Y}^l_-)_{ii}}.
\end{align}
Here $Y^l_{\pm}$ are the resummed Yukawa couplings
\begin{align*}
(\bar{Y}^l_+)_{l1} &= Y^l_{l1} + iB_{l1} - \frac{iY^l_{l2} m_{N_1} (m_{N_1} A_{12} + m_{N_2} A_{21})}{m_{N_1}^2 - m_{N_2}^2 + 2i A_{22} m_{N_1}^2}, \\
(\bar{Y}^l_+)_{l2} &= Y^l_{l2} + iB_{l2} - \frac{iY^l_{l1} m_{N_2} (m_{N_2} A_{21} + m_{N_1} A_{12})}{m_{N_2}^2 - m_{N_1}^2 + 2i A_{11} m_{N_2}^2}, \\
(\bar{Y}^l_-)_{l1} &= Y^{l*}_{l1} + iB^*_{l1} - \frac{iY^{l*}_{l2} m_{N_1} (m_{N_1} A^*_{12} + m_{N_2} A^*_{21})}{m_{N_1}^2 - m_{N_2}^2 + 2i A_{22} m_{N_1}^2}, \\
(\bar{Y}^l_-)_{l2} &= Y^{l*}_{l2} + iB^*_{l2} - \frac{iY^{l*}_{l1} m_{N_2} (m_{N_2} A^*_{21} + m_{N_1} A^*_{12})}{m_{N_2}^2 - m_{N_1}^2 + 2i A_{11} m_{N_2}^2}.
\end{align*}

These depend on the leptogenesis scale Yukawa couplings in the basis where the Majorana mass matrix is diagonal, $Y^l$, as well as contributions from the absorptive parts of the one-loop vertices and self-energies of the diagrams in Figure~\ref{fig:feynmamn}~(b)~and/or~(c). Those contributions are given by
\begin{align}
\Sigma_{ij}^{\text{abs}}(\slashed{p}) &= A_{ij} \slashed{p} P_L + A_{ij}^* \slashed{p} P_R \label{loop_and_self_energy_contributions}\\&= \sum_{l'=1}^3 \left( \frac{Y^l_{l'i} Y^{l*}_{l'j}}{16\pi} \slashed{p} P_L + \frac{Y^{l*}_{l'i} Y^l_{l'j}}{16\pi} \slashed{p} P_R \right) \notag\\
\mathcal{V}_{li}^{\text{abs}}(\slashed{p}) &= \frac{B_{li}}{\sqrt{p^2}} \slashed{p} P_L\notag \\&= - \sum_{l'=1}^3 \sum_{\substack{j=1,2 \notag\\ (j \neq i)}} \frac{Y^{l*}_{l'i} Y^l_{l'j} Y^l_{lj}}{16\pi\sqrt{p^2}} \slashed{p} P_L f\left(\frac{m_{N_j}^2}{p^2}\right)\notag \\\notag
\end{align}
Here $A_{ji} = A_{ij}^*$, $f(x) = \sqrt{x}[1 - (1+x)\ln(1+1/x)]$ is the Fukugita-Yanagida one-loop function\cite{Lepto_paper_original_Fukugita:1986hr}, $\slashed{p}$ is Feynman notation for the contraction of the four-momentum with $\gamma^\mu$, and $P_L/P_R$ are the chiral projection operators. 

The coupled Boltzmann equations describing the rate of change of the RH neutrino density $\eta_{N_i}(z)=n_{N_i}(z)/n_\gamma(z)$ and lepton asymmetry $\eta_L(z)$ as a function of $z=m_{N_1}/T$ are given by
\begin{widetext}
\begin{align}
\frac{d\eta_{N_i}}{dz} =& \frac{z}{H(z=1)} \left[ \left(1 - \frac{\eta_{N_i}}{\eta_{N_i}^{\text{eq}}}\right) \left( \Gamma^{D (i)} + \Gamma_{\text{Yukawa}}^{S (i)} + \Gamma_{\text{Gauge}}^{S (i)} \right) \right. 
\quad \left. - \frac{1}{4} \eta_L \delta_{N_i} \left( \Gamma^{D (i)} + \tilde{\Gamma}_{\text{Yukawa}}^{S (i)} + \tilde{\Gamma}_{\text{Gauge}}^{S (i)} \right) \right],\notag\\ 
\frac{d\eta_L}{dz} =& - \frac{z}{H(z=1)} \left\{ \sum_{i=1}^3 \delta_{N_i} \left(1 - \frac{\eta_{N_i}}{\eta_{N_i}^{\text{eq}}}\right) \left( \Gamma^{D (i)} + \Gamma_{\text{Yukawa}}^{S (i)} + \Gamma_{\text{Gauge}}^{S (i)} \right.\right)\quad\notag\\
& \left.\:\:\:\:\:\:\:\:\:\:\:\:\:\:\:\:\:\:\:\:\:\:\:\:\:\:\:\:\:\:\:\:\:\:\:\:\:\:\:\:\:\:\:\:\:\:\:\:\:\:\:\:\:\:\:\:\:\:\:\:\:\:\:+ \frac{1}{4} \eta_L \left[ \sum_{i=1}^3 \left( \Gamma^{D (i)}+ \Gamma_{\text{Yukawa}}^{W (i)} + \Gamma_{\text{Gauge}}^{W (i)} \right) + \Gamma_{\text{Yukawa}}^{\Delta L=2} \right] \right\} 
\end{align}
\end{widetext}
Taking $H(T)$ to be the Hubble parameter, $M_{\rm Planck} = 1.2\times10^{16}$\,TeV to be the Planck mass and $g_*\approx107$ to be the number of
relativistic degrees of freedom of the SM then
\begin{equation}
    H(T)=1.66g_*^{1/2}\dfrac{T^2}{M_{\rm Planck}},
\end{equation}
where, given the following definition for the CP-conserving collision terms 
\begin{equation}
\gamma_Y^X \equiv \gamma(X \rightarrow Y) + \gamma(\overline{X} \rightarrow \overline{Y}) ,
\end{equation}
Pilaftsis and Underwood define
\begin{align}
\Gamma^{D (i)} &= \frac{1}{n_\gamma} \gamma_{LH}^{N_i} ,\\
\Gamma_{\text{Yukawa}}^{S (i)} &= \frac{1}{n_\gamma} \left( \gamma_{Qu^C}^{N_i L} + \gamma_{LQ^C}^{N_i u^C} + \gamma_{Lu}^{N_i Q} \right) ,\notag\\
\tilde{\Gamma}_{\text{Yukawa}}^{S (i)} &= \frac{1}{n_\gamma} \left( \frac{\eta_{N_i}}{\eta_{N_i}^{\text{eq}}} \gamma_{Qu^C}^{N_i L} + \gamma_{LQ^C}^{N_i u^C} + \gamma_{Lu}^{N_i Q} \right) ,\notag\\
\Gamma_{\text{Gauge}}^{S (i)} &= \frac{1}{n_\gamma} \left( \gamma_{LH}^{N_i V_\mu} + \gamma_{H^\dagger V_\mu}^{N_i L} + \gamma_{LV_\mu}^{N_i H^\dagger} \right) ,\notag\\
\tilde{\Gamma}_{\text{Gauge}}^{S (i)} &= \frac{1}{n_\gamma} \left( \gamma_{LH}^{N_i V_\mu} + \frac{\eta_{N_i}}{\eta_{N_i}^{\text{eq}}} \gamma_{H^\dagger V_\mu}^{N_i L} + \gamma_{LV_\mu}^{N_i H^\dagger} \right) ,\notag\\
\Gamma_{\text{Yukawa}}^{W (i)} &= \frac{2}{n_\gamma} \left( \gamma_{Qu^C}^{N_i L} + \gamma_{LQ^C}^{N_i u^C} + \gamma_{Lu}^{N_i Q} + \frac{\eta_{N_i}}{2\eta_{N_i}^{\text{eq}}} \gamma_{Qu^C}^{N_i L} \right) ,\notag\\
\Gamma_{\text{Gauge}}^{W (i)} &= \frac{2}{n_\gamma} \left( \gamma_{LH}^{N_i V_\mu} + \gamma_{H^\dagger V_\mu}^{N_i L} + \gamma_{LV_\mu}^{N_i H^\dagger} + \frac{\eta_{N_i}}{2\eta_{N_i}^{\text{eq}}} \gamma_{H^\dagger V_\mu}^{N_i L} \right) ,\notag\\
\Gamma_{\text{Yukawa}}^{\Delta L=2} &= \frac{2}{n_\gamma} \left( \gamma_{L^CH^\dagger}^{\prime LH} + 2\gamma_{H^\dagger H^\dagger}^{LL} \right).\notag
\end{align}
Here $V_\mu=B_\mu,W^a_\mu$ represent the gauge bosons present above the EW symmetry breaking scale, $Q$ are the left-handed quark doublets, $u$ are the quark singlets, and $\eta^{\mathrm{eq}}_a$ are the in-equilibrium number distributions for a species $a$.

We then solve the coupled Boltzmann equations numerically in Mathematica using NDSolve. In the following section we present the results of calculations for the extreme values of the allowed parameter space of the Yukawa couplings, cutoff scale, and heavy RH neutrino mass scale. For each extreme point we also determine the minimum mass splitting needed to meet phenomenological observations. 

\subsection{Constraints on Yukawa Couplings, $\Lambda$, and Heavy RH Neutrino Mass Scale}
As mentioned previously, Eq.~(\ref{m_a_LAMBDA_y_d1_constaint_equation}), along with constraints on the individual parameters therein, substantially restrict the available parameter space. We wish to explore the boundaries of these constraints and examine the model's predictions for the BAU at each of them. 

Requiring the Yukawa coupling to be perturbatively small, we take it to fall in the region $0.1\leq y_{\mathrm{D}_1}\leq1$. Furthermore, the cutoff scale $\Lambda$ must be sufficiently high so that at the leptogenesis scale the flavons develop their VEVs --- for this reason we conservatively select the floor $\Lambda_{\mathrm{min}}=100$\,TeV. Additionally, $\Lambda$ is limited by the fact that it is inversely proportional to the Majorana mass, which is in turn constrained by the sphaleron freeze-out temperature, $T_{\rm sph}$, through
\begin{equation}
    z_{\rm crit} = \dfrac{m_{N_{i}}}{T_{\rm sph}}\approx\dfrac{m_{N_{i}}}{131\mathrm{\,GeV}},
\end{equation}
since too small a $z$ will not allow for sufficient lepton asymmetry to develop before the sphaleron process ceases \cite{Sphaleron_freeze_out_temp_Donofrio:2014pvw}. With this in mind we require $z\geq1$, which, in concert with $y_{\mathrm{D}_1}=0.1$, leads to an upper bound on the cutoff scale of $\Lambda=419$\,TeV. So, we perform calculations at the extremes of the range 
\begin{eqnarray}
0.1\:\leq  &  y_{\mathrm{D}_1}    &\leq\:1,\\
100\mathrm{\,TeV}\:\leq & \Lambda \notag&\leq\:419\mathrm{\,TeV},\notag\\\label{yukawa_and_lambda_parameter_ranges}\notag\end{eqnarray} 
which via Eqs.(\ref{majorana_mass_in_terms_of_LAMBDA_etc}) and (\ref{m_a_value}) leads to the following range on the lightest Majorana mass
\begin{equation}
    0.13\mathrm{\,TeV}\:\leq\:\:m_{N_{1}}\:\leq\:230\mathrm{\,TeV}.
\end{equation}

\subsection{Numerical Resonant Leptogenesis Results for Simple $(\sqrt{2},1;1,2)$ Alignment }
Finally, we parameterize the splitting between the degenerate Majorana masses induced by the off-diagonal $\Delta$ terms in Eq.~(\ref{diagonalized_majorana_mass_Matrix_with_mass_splitting}) like so
\begin{equation}
m_{N_1}=M-\Delta,\:\:\:\:m_{N_2}=M+\Delta=m_{N_1}(1+x_{\Delta})
\end{equation}
This parameterization is chosen to continue to mirror the formalism of Ref.~\cite{Resonant_lepto_Pilaftsis_2004}. We then take the four extreme points of $(y_{\mathrm{D}_1},\:\Lambda)$ as defined by the ranges in Eq.~(\ref{yukawa_and_lambda_parameter_ranges}) when $M\rightarrow m_{N_1}$. Then we compute for each $x_{\Delta,\mathrm{min}}$ --- the minimum mass splitting required such that the lepton asymmetry asymptotically approaches the value required to yield the observed BAU. That is, in all cases, and for both sets of initial conditions considered, the lepton asymmetry approaches a single value as $T$ increases. Given the different $z_{\mathrm{crit}}$ values for each set of parameters it may be the case that the actual lepton asymmetry produced may be higher than that of the asymptotic value. We use the asymptotic value to facilitate comparison between parameter sets. We take this value and compute the corresponding baryon asymmetry by using the sphaleron conversion factor~\cite{Harvey:1990qw,Khlebnikov:1988sr}
\begin{equation}
\eta_B=\dfrac{28}{79}\eta_L.
\end{equation}
The value $x_{\Delta,\mathrm{min}}$ is then defined as the minimum value for which $\eta_B=6.1\times10^{-10}$. Doing so in this way allows us to explore the full range of the set of parameters $(y_{\mathrm{D}_1},\:\Lambda,x_{\Delta,\mathrm{min}})$ which, in concert with the Simple $(\sqrt{2},1;1,2)$ Alignment, may yield observed BAU. 

The lepton asymmetry $\eta_L$ and heavy neutrino number density $\eta_{N_i}$ are computed using two different sets of initial conditions. First, in the ``zero initial conditions (zero IC)" set of initial conditions we assume the Majorana neutrinos start strongly out of equilibrium, and for ``thermal initial conditions (thermal IC)" we assume they begin in equilibrium. All computations correspond to the ``full" calculations presented in Ref.~\cite{Resonant_lepto_Pilaftsis_2004} in that they include the CP-violating scattering terms proportional to $\delta_{N_i}=(\Gamma^{S(i)}_{\rm Yukawa}+\Gamma^{S(i)}_{\rm Gauge})$.

In Table~\ref{tab:xcrit_results} we present the computed values of $z_{\mathrm{crit}}$ and $x_{\Delta,\mathrm{min}}$ corresponding to each of the four extreme points of $(y_{\mathrm{D}_1},\:\Lambda,x_{\Delta})$ values. From this we can see the range of viable mass splittings to be
\begin{equation}
  2.45 \times 10^{-9}\mathrm{\,GeV}\leq x_{\Delta,\mathrm{min}}\leq 4.31 \times 10^{-6}\mathrm{\,GeV}
\end{equation}
where the minimum value ensures that all points within the allowed $(y_{\mathrm{D}_1},\:\Lambda)$ parameter space will yield a sufficiently large BAU whereas the maximum allows only the $(y_{\mathrm{D}_1}=1,\:\Lambda=100\mathrm{\,TeV})$ scenario to be viable. 

\begin{table}[htbp]
    \centering
    \small
    \setlength{\tabcolsep}{3pt}
    \renewcommand{\arraystretch}{1.5}
    \begin{tabular}{|l|c|c|c|c|c|}
        \hline
        Set &$y_{\mathrm{D}_1}$ & $\Lambda$\,[TeV] & $m_{N_1}$\,[TeV] & $z_{\mathrm{crit}}$ & $x_{\Delta,\mathrm{min}}\mathrm{\,[GeV]}$ \\
        \hline
        1&0.1 & 419 & 0.13 & 1 & $2.45 \times 10^{-9}$ \\
        \hline
       2& 0.1 & 100 & 2.3 & 17.56 & $4.31 \times 10^{-8}$ \\
        \hline
        3&1 & 419 & 2.3 & 100 & $2.45 \times 10^{-7}$ \\
        \hline
       4& 1 & 100 & 230 & 1756 & $4.31 \times 10^{-6}$ \\
        \hline
    \end{tabular}
    \caption{The four extreme points of the parameters $(y_{\mathrm{D}_1},\:\Lambda)$ and their corresponding $m_{N_1}$, $z_{\rm crit}$, and $x_{\Delta,\mathrm{min}}$.}
    \label{tab:xcrit_results}
\end{table}

\begin{figure}[htbp]
    \centering
    \includegraphics[width=1\linewidth]{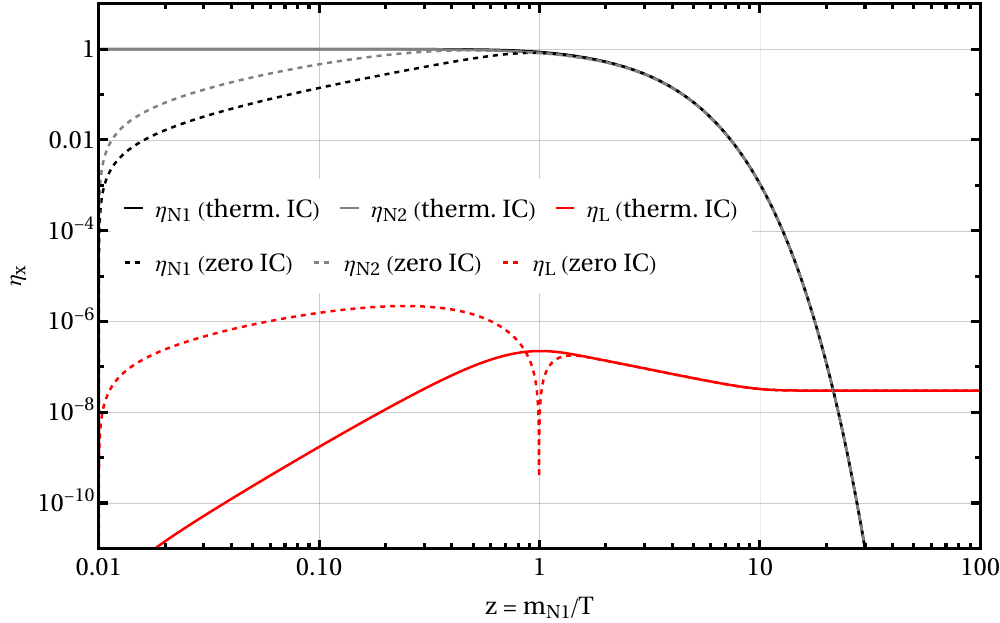}
    \caption{$\eta_L$ and $\eta_{N_i}$ for $y_{\mathrm{D}_1}=0.1,\:\:\Lambda=100\mathrm{\,TeV},\:\:x_{\Delta}=2.45\times10^{-9}\,$GeV}
    \label{fig:y_0_1___L_1___2_45e-9.png}
\end{figure}

\begin{figure}
    \centering
    \includegraphics[width=1\linewidth]{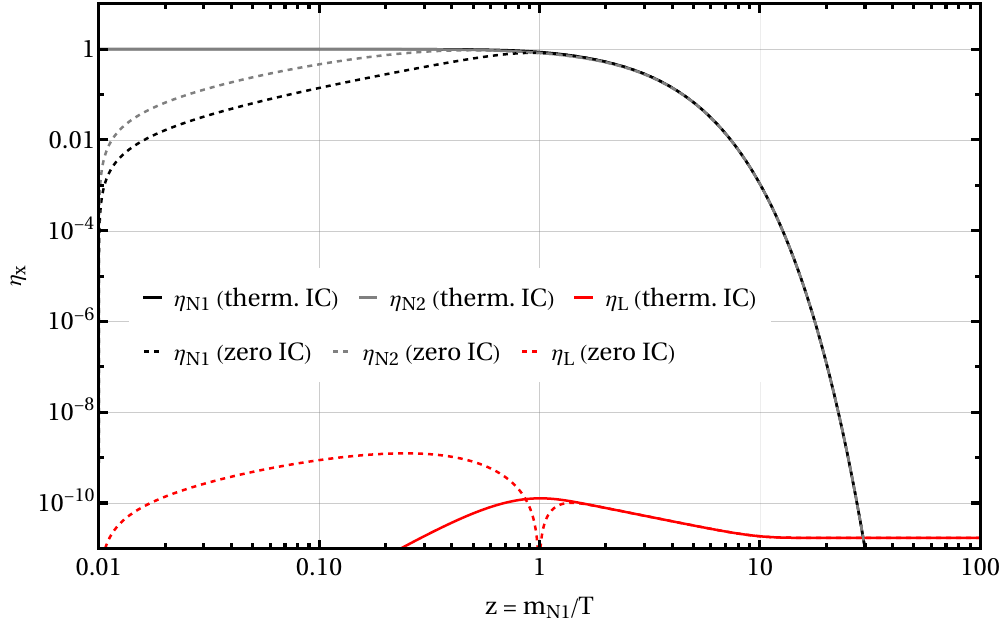}
    \caption{$\eta_L$ and $\eta_{N_i}$ for $y_{\mathrm{D}_1}=0.1,\:\:\Lambda=100\,\mathrm{TeV},\:\:x_{\Delta}=4.31\times10^{-6}\,$GeV}
    \label{fig:y_0_1___L_1___4_31e-6.png}
\end{figure}

\begin{figure}[htbp]
    \centering
    \includegraphics[width=1\linewidth]{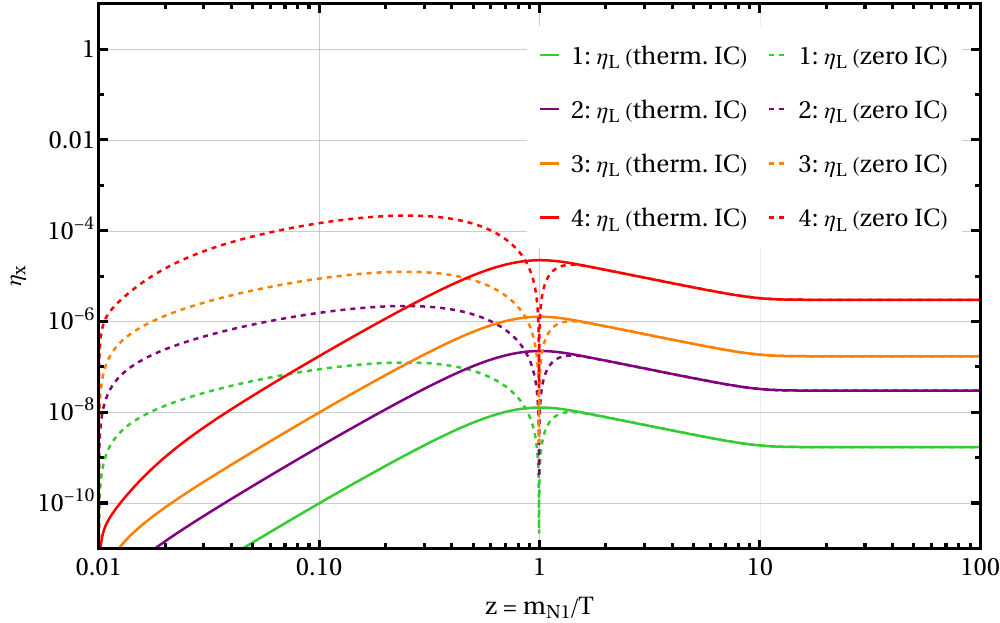}
      \caption{$\eta_L$ for $ x_{\Delta}=2.45\times10^{-9}\,$GeV for the following extreme point sets: $1(y_{\mathrm{D}_1}=0.1,\Lambda=419\mathrm{\,TeV})$, $2(y_{\mathrm{D}_1}=0.1,\Lambda=100\mathrm{\,TeV})$, $3(y_{\mathrm{D}_1}=1,\Lambda=419\mathrm{\,TeV})$, $4(y_{\mathrm{D}_1}=1,\Lambda=100\mathrm{\,TeV})$}
    \label{fig:all_parameters_all_x_are_245e-9.png}
\end{figure}

\begin{figure}[htbp]
    \centering
    \includegraphics[width=1\linewidth]{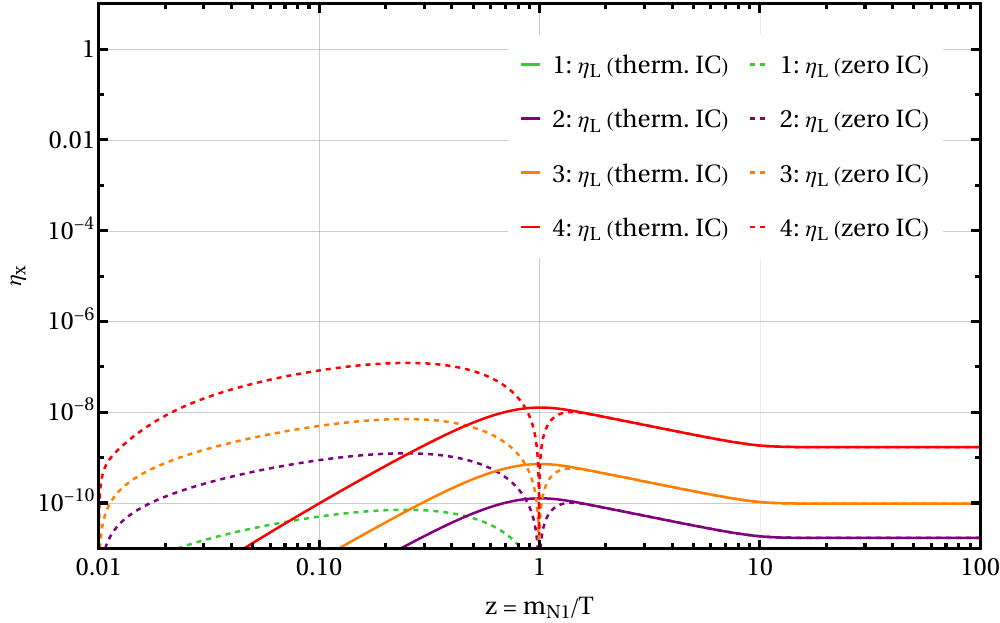}
   \caption{$\eta_L$ for $x_{\Delta}=4.36\times10^{-6}\,$GeV for the following extreme point sets: $1(y_{\mathrm{D}_1}=0.1,\Lambda=419\mathrm{\,TeV})$, $2(y_{\mathrm{D}_1}=0.1,\Lambda=100\mathrm{\,TeV})$, $3(y_{\mathrm{D}_1}=1,\Lambda=419\mathrm{\,TeV})$, $4(y_{\mathrm{D}_1}=1,\Lambda=100\mathrm{\,TeV})$}
    \label{fig:all_parameters_all_x_are_431e-6.png}
\end{figure}

Taking the set $(y_{\mathrm{D}_1}=0.1,\:\Lambda=100\mathrm{\,TeV})$ we plotted the lepton asymmetry and RH neutrino abundances as a function of $z$ under both sets of initial conditions. This is shown for the minimum value of $x_{\Delta,\mathrm{min}}=2.45 \times 10^{-9}\mathrm{\,GeV}$ in Figure~\ref{fig:y_0_1___L_1___2_45e-9.png} and for the maximum value $x_{\Delta,\mathrm{min}}=4.31 \times 10^{-6}\mathrm{\,GeV}$ in Figure~\ref{fig:y_0_1___L_1___4_31e-6.png}. Next we plotted the lepton asymmetry for all four sets of extreme points $(y_{\mathrm{D}_1},\:\Lambda)$ and both sets of initial conditions using $x_{\Delta,\mathrm{min}}=2.45 \times 10^{-9}\mathrm{\,GeV}$ in Figure~\ref{fig:all_parameters_all_x_are_245e-9.png} and $x_{\Delta,\mathrm{min}}=4.31 \times 10^{-6}\mathrm{\,GeV}$ in Figure~\ref{fig:all_parameters_all_x_are_431e-6.png}. 

In these plots we have not shown $\eta_{N_i}$ as it is essentially the same across all sets $(y_{\mathrm{D}_1},\:\Lambda,x_{\Delta,\mathrm{min}})$. 
From Figures~\ref{fig:y_0_1___L_1___2_45e-9.png}-\ref{fig:all_parameters_all_x_are_431e-6.png} and Table~\ref{tab:xcrit_results} it can be seen that the relationship between the magnitude of the large-$z$ asymptotic lepton asymmetry and $x_{\Delta}$ is essentially linear. Furthermore, aside from the magnitude, the relationship between $\eta_x$ and $z$ is essentially the same in all cases --- though it should be noted that at the same $z$ each parameter set corresponds to a different temperature $T$, since each triple $(y_{\mathrm{D}_1},\Lambda,x_{\Delta,\mathrm{min}})$ corresponds to a different $m_{N_{1}}$.

These plots demonstrate that the Simple $(\sqrt{2},1;1,2)$ Alignment contains sufficient CP asymmetry to produce enough lepton asymmetry in the early universe to in turn yield the observed BAU. Furthermore, we determined the range of $x_{\Delta,\mathrm{min}}$ that allows for sufficient resonant enhancement of the CP-violating RH neutrino decays within the allowed regions of parameter space.

\section{Conclusion}

The primary goal in this work has been to show how phenomenological model construction techniques that assume the limit of SD may be made applicable when some of the RH neutrino masses are degenerate --- an approach we named RSD.

As a first step, we introduced a new phenomenologically viable PMNS matrix ansatz that we refer to as TBC3, along the lines of TBC1 and TBC2 of Ref.~\cite{King_nice_Vev_search_Master_formula_paper_king2013minimalpredictiveseesawmodel}. The TBC3 ansatz achieves $1\sigma$ agreement with current NuFIT data and reduces the number of free parameters by assuming equality of the deviation from TBM of the solar and atmospheric angles, all while employing only simple multiples of the Cabibbo parameter.

In RSD the hierarchy of the left-handed neutrino masses is generated, not through the RH neutrino mass hierarchy such as in SD, but instead purely through the Yukawa couplings. Within the RSD framework, and neglecting the third ultra-heavy RH neutrino $N_3$, we used the Master Formula of Ref.~\cite{King_nice_Vev_search_Master_formula_paper_king2013minimalpredictiveseesawmodel} along with $\chi^2$ analysis to identify possible combinations of the Dirac neutrino mass matrix elements that may yield correct low-energy observables. In particular, we searched for simple values. The most interesting set of values we found is the Simple $(\sqrt{2},1;1,2)$ Alignment which produces $1\sigma$ agreement with all mixing angles, the ratio of the two non-zero neutrino mass eigenvalues, and the CP-violating Dirac phase. 

Furthermore, we constructed an explicit $A_4\times(Z_3)^4\times(Z_2)^2$-symmetric lepton mass Lagrangian compatible with both RSD and resonant leptogenesis. The resulting Lagrangian extends the SM field content by adding three RH neutrinos, $N_i$, four scalar flavons $\phi_i$ and $\varphi$, and the three scalar fields $\chi_i$. We then extended the SM symmetry by $A_4\times(Z_3)^4\times(Z_2)^2$ which, via careful charge assignment, produces a Lagrangian which, up to terms of dimension five, has the exact desired lepton sector structure to be compatible with RSD and resonant leptogenesis. 

We also established the connection between the parameters of the low-energy RSD model used to search for viable Dirac mass matrix elements and the underlying Lagrangian. From this we derived constraints on the heavy Majorana mass scale --- which also depends on the choice of high-energy cutoff scale $\Lambda$ and the single independent Yukawa coupling $y_{\mathrm{D}_1}$. For the numerical calculations we proceeded with deriving the constraints based on the Simple $(\sqrt{2},1;1,2)$ Alignment. We numerically solved the coupled Boltzmann equations and computed the predictions of the model for lepton asymmetry as a function of $z=m_{N_1}/T$. This was done at the four extreme points of $(y_{\mathrm{D}_1},\:\Lambda)$. For each extreme point we then computed the minimum size of the Majorana mass splitting needed in order to generate sufficient lepton asymmetry. In doing so, we defined the boundaries of the parameter space of the three quantities $(y_{\mathrm{D}_1},\:\Lambda,x_{\Delta,\mathrm{min}})$ for which the observed BAU may be produced using the Simple $(\sqrt{2},1;1,2)$ Alignment.

From the leptogenesis computations, using the Simple $(\sqrt{2},1;1,2)$ Alignment, we determined the viable ranges of $(y_{\mathrm{D}_1},\:\Lambda,x_{\Delta,\mathrm{min}})$ to be 
\begin{eqnarray}
0.1\:\leq  &  y_{\mathrm{D}_1}    &\leq\:1,\\
100\mathrm{\,TeV}\:\leq & \Lambda &\leq\:419\mathrm{\,TeV},\notag\\
2.45 \times 10^{-9}\mathrm{\,GeV}\:\leq&x_{\Delta,\mathrm{min}}&\leq\:4.31 \times 10^{-6}\mathrm{\,GeV}.\notag
\end{eqnarray}
It was further demonstrated that the above range of parameters corresponds to the following mass range for the lightest Majorana neutrino
\begin{equation}
0.13\mathrm{\,TeV}\:\leq\:\:m_{N_{1}}\:\leq\:230\mathrm{\,TeV}.   
\end{equation}
Thus we have demonstrated that within the allowed parameter space there is a range of possible mass splittings that can produce sufficient resonant enhancement of the CP-violating processes that underpin leptogenesis and lead to baryogenesis. 

We have taken an indirect model-building approach and therefore did not propose a symmetry which may naturally produce the Simple $(\sqrt{2},1;1,2)$ Alignment. Furthermore, we have not suggested more fundamental origins for the TBC3 ansatz or a particular origin for the RH neutrino mass splittings. These comprise interesting avenues for future research. Finally, we note that the lower end of the allowed mass range of $m_{N_1}$ lies within the energy reach of the LHC. However, the collider phenomenology depends on the detailed properties of the RH neutrino, which are beyond the scope of the present work.

\bibliographystyle{apsrev4-2}  
\bibliography{bibo}
\end{document}